\documentclass[11pt,a4paper]{article}
\usepackage{jheppub}

\usepackage{amsmath,amssymb,mathtools,mleftright,orcidlink,amsthm,booktabs,mathdots,cancel}
\usepackage{CJKutf8}
\usepackage[german,french,latin,british]{babel}

\usepackage{tikz-cd}
\usetikzlibrary{babel}
\usepackage{cleveref}

\title{Symmetries Beget Symmetries: Ghostly Higher-Form Symmetries and the Descent Equation}
\author[a,b]{Leron Borsten\textsuperscript{\orcidlink{0000-0001-9008-7725}},}
\author[a]{Dimitri Kanakaris\textsuperscript{\orcidlink{0009-0001-7716-851X}},}
\author[a]{and Hyungrok Kim (\begin{CJK*}{UTF8}{bsmi}金炯錄\end{CJK*})\textsuperscript{\orcidlink{0000-0001-7909-4510}}}
\affiliation[a]{Department of Physics, Astronomy and Mathematics, University of Hertfordshire\\ Hatfield, Hertfordshire AL10 9AB, United Kingdom}
\affiliation[b]{Blackett Laboratory, Imperial College London\\ London SW7 2AZ, United Kingdom}
\emailAdd{l.borsten@herts.ac.uk}
\emailAdd{d.kanakaris-decavel@herts.ac.uk}
\emailAdd{h.kim2@herts.ac.uk}
\abstract{
Viewed through the lens of the Batalin--Vilkovisky formalism, we demonstrate that higher-form 
currents with nonzero ghost number also
 define higher-form symmetries, directly analogous  to the standard higher-form symmetries  with ghost number zero.
These ghostly higher symmetries descend from and into conventional  higher-form symmetries
via chains of descent equations familiar from the theory of anomalies and topological field theories.
We give examples of such chains of ghostly symmetries in Maxwell theory, Abelian and non-Abelian higher gauge theory, Yang--Mills theory, and beyond.
}
\keywords{Higher-form symmetry, Batalin--vilkovisky formalism, higher gauge theory, topological field theory}
\arxivnumber{2509.15978}

\begin{document}

\maketitle

\newcommand{\be}{\begin{equation}}
\newcommand{\ee}{\end{equation}}
\newcommand{\on}{\operatorname}
\newcommand{\vOmega}{{\check\Omega}}

\section{Introduction and Summary}

We bring together generalised global symmetries and the Batalin--Vilkovisky formalism  to demonstrate that global higher-form (or $p$-form) symmetries yield families of (ghostly) higher-form symmetries via the descent equations familiar from topological quantum field theory (TQFT) and anomalies. 

Generalised symmetries were introduced in the paradigm-shifting work of \cite{Gaiotto:2014kfa} and have since led to manifold advances in a variety of domains. See the reviews \cite{Cordova:2022ruw,Brennan:2023mmt,Gomes:2023ahz,Bhardwaj:2023kri,Luo:2023ive}. The global  higher-form symmetries, in particular, provide  prominent examples that  have appeared previously in a variety of important contexts, ranging from confinement to  symmetry protected topological phases \cite{Polyakov:1976fu,Polyakov:1975rs,tHooft:1977nqb,Kovner:1992pu,deWildPropitius:1995hk,Alford:1990fc,Bucher:1991bc,  Pantev:2005rh,Pantev:2005zs,Hellerman:2006zs,Nussinov:2009zz,Kapustin:2013uxa}. The basic intuition  underpinning $p$-form symmetries is  immediate in the context of Abelian continuous symmetries\footnote{Of course, this is merely the tip of a symmetry iceberg, but serves to build the intuition and constitutes our focus.}. In $d$ spacetime dimensions, a conventional ($0$-form) global symmetry, denoted $G^{[0]}$,  is generated by a conserved  $(d-1)$-form Noether current $J_{d-1}$\footnote{Note that we adopt the convention used in the generalised symmetries literature of calling the closed  \((d-p-1)\)-form $J$ the current (as opposed to the convention of regarding a current as  a divergence free $(p+1)$-multivector density).} . So,    a global $p$-form symmetry, denoted $G^{[p]}$,  ought be generated by a conserved $(d-p-1)$-form Noether current $J_{d-p-1}$, of course. When considering gauge theories in general, and  the Ward identities in particular,  a natural question arises: what is the fate of the higher-form symmetries in the Batalin--Vilkovisky framework?

The Batalin--Vilkovisky formalism \cite{Batalin:1977pb,Batalin:1981jr,Batalin:1983ggl,Batalin:1984ss,Batalin:1985qj}
offers a powerful and elegant method of dealing with gauge theories, including those with open gauge algebras in a wide variety of settings, as reviewed in \cite{Henneaux:1989jq,Henneaux:1992ig,Gomis:1994he,Barnich:2000zw,Fiorenza:2004sg,Fuster:2005eg,Qiu:2011qr,Mnev:2017oko,Barnich:2018gdh,Cattaneo:2019jpn,mnev2019,Cattaneo:2023hxv}. The basic idea is to provide an algebraic characterisation, more amenable to path-integral quantisation, of the space of classical solutions modulo gauge transformations by introducing ghosts and antifields. The Batalin--Vilkovisky operator  relates physical fields to  ghosts and, so,  has potential implications for higher-form symmetries.   Indeed, aspects of higher-form symmetries in the context of the Batalin--Vilkovisky formalism have been  considered before they were so named \cite{Brandt:1997cz,Brandt:1996nr,Brandt:1996uv}, as well as more recently in \cite{Dneprov:2025eoi, Borsten:2025pvq}. 
Ordinarily, higher-form symmetries are associated to Noether currents that have zero ghost number.
However, from the perspective of the Batalin--Vilkovisky formalism, it is very unnatural to restrict to zero ghost number;
one should treat all fields (including Noether currents) of all ghost numbers uniformly.
Thus we are led to postulate the existence of \emph{ghostly} \(p\)-form symmetries \(G^{[p,q]}\), generalising the conventional \(p\)-form symmetries \(G^{[p]}\), whose associated Noether currents have nonzero ghost numbers \(q\). These naturally act on Wilson and 't~Hooft operators with nonzero ghost number.

How are we to find such ghostly symmetries, and how are they related to non-ghostly symmetries?
One answer lies in the \emph{descent equations} that are foundational in topological field theories \cite{Witten:1988ze} (reviewed in \cite{Labastida:1997pb,Labastida:2005zz,Chen:2008yp}) and anomalies \cite{Stora:1976kd,Stora:1983ct,Zumino:1983ew,Zumino:1983rz,Zumino:1984ws} (reviewed in \cite{Bertlmann:1996xk,Harvey:2005it,Bilal:2008qx}).
More recently, the decent equations have appeared in constructions of symmetry topological field theories (SymTFTs) via symmetry inflow \cite{Gagliano:2024off}.
In topological field theories, starting from a \(p\)-form observable with ghost number \(q\), one uses the descent equation to find another observable, this time with form degree \(p+1\) and ghost number \(q-1\); crucially, if one starts with a topological observable, its descendants are also guaranteed to be topological.
This property is precisely what is needed in the construction of higher-form symmetries, which are precisely topological operators inserted along submanifolds.
Therefore, we find that the descent equations also relate higher-form symmetries of different ghost numbers amongst each other: there exists a canonical span of group homomorphisms
\begin{equation}
    G^{[p,q]}\leftarrow G^{[p,q];[p+1,q-1]}\to G^{[p+1,q-1]}
\end{equation}
defined by the descent equation, where the group \(G^{[p,q];[p+1,q-1]}\) parameterises the solutions of the descent equation. Thus, starting from a known higher-form symmetry of ghost number zero, one can take \emph{descendants} to find more higher-form symmetries of negative ghost number: in short, \emph{symmetries beget symmetries}.

This paper is organised as follows.
In \cref{sec:BV}, we briefly review the aspects of the Batalin--Vilkovisky formalism required for the treatment of continuous and discrete global higher-form symmetries and their (ghostly) descendants. In \cref{sec:general_discussion}, we  present the general formalism of how higher-form symmetries and their higher-form-symmetry progeny arise in the Batalin--Vilkovisky formalism via the   descent equation.
Subsequent sections present a series of examples in increasing order of difficulty:
\cref{sec:free_matter} discusses the trivial case of descent into the antifield \((-1)\)-form symmetry that occurs in any free field theory;
\cref{sec:abelian} discusses the descendants of the electric and magnetic symmetries in an Abelian gauge theory (including Maxwell theory);
\cref{sec:tqft} reinterprets the familiar observables of cohomological (or Witten-type) topological field theories as ghostly higher symmetries;
\cref{sec:centre} discusses the descendants of the centre one-form symmetry in Yang--Mills theory;
finally, \cref{sec:higher_gauge} discusses the higher analogues of centre symmetry in adjusted higher gauge theory following the discussion of \cite{Borsten:2025diy}.

\section{The Batalin--Vilkovisky formalism revisited}\label{sec:BV}

The Batalin--Vilkovisky formalism describes a physical theory with gauge symmetry (such as continuum and lattice quantum field theories) in the language of a differential graded manifold \(X\).
That is, a `manifold' with $\mathbb Z$-graded coordinates equipped  with a ghost number \(+1\) differential
\begin{equation}
    Q\colon\mathcal C^\infty(X)\to\mathcal C^\infty(X), \qquad Q^2=0
\end{equation}
on the algebra of smooth functionals on \(X\).\footnote{This differential graded manifold \(X\) comes with a symplectic structure as well as a density, which then equips the ring of functions on \(X\) with the structure of a Batalin--Vilkovisky algebra. This is not important for what follows, however.} The graded manifold  \(X\)  should be thought of as the space of all (off-shell) fields of all ghost numbers including the physical fields (that carry ghost number \(0\)), the  ghosts (that carry ghost number \(+1, +2,\ldots\)) and antifields (that carry ghost number \(-1,-2,\ldots\)), while $Q$ is the (classical) BV operator that encodes, amongst other things, the gauge transformations. For more details, the reader is referred to \cite{Henneaux:1992ig, Jurco:2018sby} and references therein. Here we focus on formulating local  operators in the Batalin--Vilkovisky formalism. We begin with the general formalism in \cref{ssec:BVops}, then consider $\mathbb R$-valued differential-form operators on smooth spacetimes $M$ in \cref{ssec:BVopsM}, and finally we generalise to Abelian group $G$-valued differential-form operators on `lattice spacetimes'  $\Lambda$ in \cref{ssec:discrete_differential_form} following \cite{Borsten:2025diy}. The latter is required to accommodate discrete higher global symmetries, as well as lattice models.  

\subsection{Local operators in the Batalin--Vilkovisky formalism}\label{ssec:BVops}

For a set $Y$, a (possibly nonlocal) $Y$-valued operator of the system is given by a smooth function \(\phi\in\mathcal C^\infty(X, Y)\) on \(X\) with codomain $Y$. When we leave the codomain unspecified, we assume it is $\mathbb R$. Since the codomain  is always regarded as being ungraded,  the functions carry  ghost number dual to $X$. When a function carries homogeneous degree $q$  we say it has ghost number $q$.

For example, if we have the theory of a real-valued scalar field on spacetime \(M\), then (the graded manifold underlying) \(X\) is given by the function space
\begin{equation}
    X = \mathcal C^\infty(M,\mathbb R)\times\mathcal C^\infty(M,\mathbb R)[-1] = \mathcal C^\infty(M,\mathbb R\times\mathbb R[-1]),
\end{equation}
where the first factor \( \mathcal C^\infty(M,\mathbb R)\) corresponds to the scalar fields and the second factor \( \mathcal C^\infty(M,\mathbb R)[-1]\)\footnote{Here, $V[k]$ denotes the ghost number degree shift of $V$ by $-k$. Thus, in this example, the $C^\infty(M,\mathbb R)[-1]$ factor of $X$ has degree $+1$ so that $\phi^+: C^\infty(M,\mathbb R)[-1]\to \mathbb R$ has ghost degree $-1$. That is, the shift $V[k]$ induces a shift of $k$ for the functions on $V$.} corresponds to  their   antifields. For each spacetime point \(x\in M\), then, we have the coordinate functions
\begin{align}
\phi(x)&\colon
\begin{aligned}
    X&\to\mathbb R\\
    (f,g)&\mapsto f(x),
\end{aligned}
&
\phi^+(x)&\colon
\begin{aligned}
X&\to\mathbb R\\
(f,g)&\mapsto g(x),
\end{aligned}
\end{align}
which corresponds to the scalar field and antifield local operators \(\phi(x)\) and \(\phi^+(x)\), respectively. For a free theory with kinetic term given by $\phi \mathrm {D} \phi$ for $\mathrm {D}$ some differential operator (typically $\square - m^2$) then the action of $Q: \mathcal C^\infty(X)\to \mathcal C^\infty(X)$ would be given by 
\begin{equation}
	Q \phi^+ = \mathrm D \phi, \quad Q \phi =0,
\end{equation}
where
\begin{equation}
\begin{array}{lllll}
 \mathrm D \phi &\colon &  X&\to &\mathbb R\\
& 	&  (f,g)&\mapsto &\mathrm D f (x).
\end{array}
\end{equation}

Similarly, for each vector \(V\in\mathrm T_xM\), we have the vector field local operator
\begin{equation}
\partial_V\phi(x)\colon
\begin{aligned}
    X&\to\mathbb R\\
    (f,g)&\mapsto V[f](x).
\end{aligned}
\end{equation}
On the other hand, \(\mathcal C^\infty(X)\) contains nonlocal operators as well, such as
\begin{equation}
\phi(x)\phi(y)\colon
\begin{aligned}
    X&\to\mathbb R\\
    (f,g)&\mapsto f(x)f(y)
\end{aligned}
\end{equation}
for some fixed distinct \(x,y\in M\).\footnote{To avoid technicalities and for generality, we are deliberately leaving the notion of `local' operator informal. For rigorous approaches, see e.g.\ \cite{Costello:2021jvx,Costello:2016vjw,Cattaneo:2011syo,Cattaneo:2012qu,Cattaneo:2015vsa}.}

\subsubsection{Differential-form operators on a smooth spacetime}\label{ssec:BVopsM}
The above example of scalar field operators living on a smooth spacetime $M$ generalises straightforwardly to differential-form operators, as we briefly review here.

Suppose that \(M\) is a smooth manifold. In the Batalin--Vilkovisky formalism, then, a (possibly composite) scalar (zero-form) field \(\alpha\) corresponds to a smooth assignment
\begin{equation}
    \alpha\colon M\to \mathcal C^\infty(X)
\end{equation}
from a spacetime point \(x\in M\) to an element \(\alpha(x)\in\mathcal C^\infty(X)\) that corresponds to a local operator at \(x\); more generally, a \(p\)-form field corresponds to a smooth assignment
\begin{equation}\label{eq:BVpform}
    \alpha\colon \mathrm T^{\wedge p}M\to \mathcal C^\infty(X)
\end{equation}
that maps  a spacetime point \(x\in M\) and a \(p\)-multivector \(V\in \mathrm T_x^{\wedge p}M\) to a local operator \(\alpha(V, x)\in\mathcal C^\infty(X)\)  linearly with respect to \(V\). In a coordinate basis $\{\partial_\mu\}$ for $\mathrm T M$, we may write $\alpha(x) = \alpha_{\mu_1\dotso\mu_p}(x)\mathrm dx^{\mu_1}\wedge \cdots \wedge dx^{\mu_p}$ and  \(\alpha(V, x) = V^{\mu_1\dotso\mu_p}\alpha_{\mu_1\dotso\mu_p}(x)\), where $\alpha_{\mu_1\dotso\mu_p}(x)\in C^\infty(X)$.

As a result, we obtain an algebra \(\mathcal A\) of local differential form operators. This algebra is bigraded according to form degree \(p\in\{0,\dotsc,\dim M\}\) and ghost number \(q\in\mathbb Z\):
\begin{equation}
    \mathcal A=\bigoplus_{p,q}\mathcal A^{p,q}.
\end{equation}
The algebra \(\mathcal A\) is closed under the wedge product, which is bilinear and under which the bidegrees are additive:
\begin{equation}
    \wedge\colon\mathcal A^{p,q}\otimes\mathcal A^{p',q'}\to\mathcal A^{p+p',q+q'}.
\end{equation}
The algebra \(\mathcal A\) also admits the Hodge star, which preserves ghost number but changes the form degree:
\begin{equation}
    \star\colon\mathcal A^{p,q}\to\mathcal A^{\dim(M)-p,q}.
\end{equation}
Finally, the algebra \(\mathcal A\) comes with two different nilquadratic derivations, namely the de~Rham differential \(\mathrm d\) and the Batalin--Vilkovisky differential \(Q\):
\begin{align}
    \mathrm d\colon\mathcal A^{p,q}&\to\mathcal A^{p+1,q},&
    Q\colon\mathcal A^{p,q}&\to\mathcal A^{p,q+1}.    
\end{align}
Both operators obey the graded Leibniz rule with respect to the exterior product:
\begin{align}
    \mathrm d(\alpha\wedge\beta)&=\mathrm d\alpha\wedge\beta+(-1)^p\alpha\wedge\mathrm d\beta\\
    Q(\alpha\wedge\beta)&=Q\alpha\wedge\beta+(-1)^q\alpha\wedge Q\beta
\end{align}
for \(\alpha\in\mathcal A^{p,q}\) and \(\beta\in\mathcal A^{r,s}\). Furthermore, the two differentials commute:\footnote{
    This is because our sign conventions are such that \(\alpha\beta=(-1)^{pp'+qq'}\beta\alpha\) for \(\alpha\in\mathcal A^{p,q}\) and \(\beta\in\mathcal A^{p',q'}\). A change of sign conventions (to \(\alpha\beta=(-1)^{(p+q)(p'+q')}\beta\alpha\) for instance) makes \(\mathrm d\) and \(Q\) anticommute instead; this is merely a sign convention and leads to equivalent mathematics and physics \cite[p.~62]{zbMATH01735158}.
}
\begin{equation}
    \mathrm dQ=Q\mathrm d.
\end{equation}

\subsubsection{Differential-form operators with discrete coefficients}\label{ssec:discrete_differential_form}
The previous subsection constructed the bigraded algebra \(\mathcal A\) of differential-form operators, which suffice to discuss higher-form symmetries that are continuous and, hence, where the corresponding Noether currents are ordinary differential forms. However, this does not yet suffice for  \emph{discrete} higher-form symmetries, fundamentally because the de~Rham cohomology of differential forms can only deal with real coefficients. Instead, in the discrete case one must use another cohomology theory, such as cellular cohomology, simplicial cohomology, or Čech cohomology, which work over discrete coefficients and furnish a notion of discrete-valued cochains that serve as analogues of differential forms.\footnote{Note, however, that unlike the wedge product of ordinary differential forms (which is graded-commutative before passing to cohomology), the cup products of these cochains are not in general graded-commutative before passing to cohomology. (The cup product on cohomology classes is, of course, graded-commutative.)}

For simplicity of exposition, let us initially ignore the Batalin--Vilkovisky structures  and work with cellular cohomology\footnote{Later, in \cref{sec:centre}, we will work with Čech cohomology for gauge theory, which is more convenient in that context; the story here can be translated into the Čech language straightforwardly.}, which is familiar  in the context of  lattice theories (see e.g.\ \cite{Borsten:2025diy}). Indeed, standard lattice gauge theory is a helpful model to have in mind, for those so oriented.
Analogously, suppose that \(M\) is a manifold equipped with the structure of a tiling \(\Lambda\), such that \(M\) is realised as a cellular complex (or CW-complex).
In this case, cellular cochains \cite{hatcher} serve as a discrete analogue of differential forms \cite{Borsten:2025diy}. Concretely, let \(\Lambda_p\) be the set of \(p\)-dimensional cells of the cellular structure for \(p\in\{0,\dotsc,\dim M\}\). These sit in analogy to the set of $p$-multivectors in the smooth case, $\mathrm T^{\wedge p}M$. Then a  \(p\)-form (or cellular cocycle) with coefficients in an Abelian group \(G\) is, analogously, a mapping \(\alpha\colon \Lambda_p\to G\).

Let us denote, as usual, the space of such $G$-valued  \(p\)-forms \(\Omega^p(\Lambda;G)\). Then there exist differentials,
\begin{equation}
    \mathrm d\colon\Omega^p(\Lambda;G)\to\Omega^{p+1}(\Lambda;G), \qquad \mathrm d^2=0,
\end{equation}
induced by the cocycle coboundary operators,  which  define a cohomology that agrees with the ordinary singular cohomology with coefficients in \(G\).

Cellular cochains admit a Hodge dual as follows. Tilings enjoy a duality generalising that of planar graphs and polytopes so that there is a dual tiling \(\hat\Lambda\), with a canonical bijection
\begin{equation}
    \star\colon\Lambda_p\to\hat\Lambda_{\dim(M)-p},
\end{equation}
that induces a Hodge dual map
\begin{equation}
      \star\colon \Omega^p(\Lambda;G)\to\Omega^{\dim(M)-p}(\hat\Lambda;G).
\end{equation}
Cellular cochains in general do not admit a wedge product except in the case where the cellular structure is in fact a triangulation. In that case, one can define a cup product of cochains:
\begin{equation}
    \Omega^p(\Lambda;G)\otimes_G\Omega^q(\Lambda;G)\to\Omega^{p+q}(\Lambda;G).
\end{equation}
The dual of a triangulation is, however, rarely a triangulation. One can always pass to a more refined cellular structure (which can be chosen to be a triangulation); if \(\Lambda'\) refines \(\Lambda\) (such that there is a cellular morphism \(\Lambda\to\Lambda'\)), then there exists a canonical map
\begin{equation}
    \Omega^p(\Lambda';G)\to\Omega^p(\Lambda;G).
\end{equation}
The above map defines a quasi-isomorphism between the corresponding cellular cochain complexes.
In this way, one obtains a workable theory of differential forms with coefficients in an Abelian group \(G\).

Comparing with \cref{ssec:BVopsM}, the generalisation of the above theory of ghost number zero discrete $p$-form operators to the graded Batalin--Vilkovisky setting is evident. Following the discussion in \cref{ssec:BVopsM}, in particular \eqref{eq:BVpform}, given a differential graded manifold $X$ we may analogously define a \(p\)-form operator with coefficients in \(G\) with respect to a cellular structure \(\Lambda\) to be a mapping
\begin{equation}
  \alpha\colon  \Lambda_p\to \mathcal C^\infty(X,G)
\end{equation}
that is local in a suitable lattice sense. So, for instance, a zero-form operator associates an element of \(\mathcal C^\infty(X,G)\) to each lattice site \(x\in\Lambda_0\), and a one-form operator associates an element of \(\mathcal C^\infty(X,G)\) to each link (edge) \(e\in\Lambda_1\).
The collection of differential form operators with coefficients in \(G\), which we denote as \(\mathcal A(\Lambda,G)\), is similarly bigraded:
\begin{equation}
    \mathcal A(\Lambda;G)=\bigoplus_{p,q}\mathcal A^{p,q}(\Lambda;G).
\end{equation}
On it we have the operators \(\mathrm d\) and \(Q\), of degrees \((1,0)\) and \((0,1)\) respectively; if \(\Lambda\) is a triangulation, then \(\mathrm d\) and \(Q\) obey the Leibniz rule as in the continuum case.

\section{Higher form symmetries and their descendants}\label{sec:general_discussion}

\subsection{Symmetries and the descent equation}
A \(p\)-form symmetry is a topological invertible codimension \(p+1\) operator.
In the case of a \(\operatorname U(1)\)-valued symmetry,
we may write such operators \(U(\Sigma_{d-p-1})\) on a submanifold \(\Sigma_{d-p-1}\) of codimension \(p+1\) as
\begin{equation}\label{eq:psym}
    U(\Sigma_{d-p-1})
    = \exp\mleft(\mathrm i\int_{\Sigma_{d-p-1}}\alpha J\mright)
\end{equation}
for some \((d-p-1)\)-form current \(J\) and parameter \(\alpha\in\mathbb R\). We require this operator to be invariant under continuous deformations of \(\Sigma_{d-p-1}\); a sufficient condition for this to hold is that \(\mathrm dJ\overset{\textsc{eom}}=0\), where \(\overset{\textsc{eom}}=\) denotes equality on shell (that is, up to equations of motion).

However, one of the key lessons of the Batalin--Vilkovisky formalism is that the concept of `on-shellness' is resolved\footnote{We use  `resolved' loosely here   in the technical sense that goes back, at least, to  Hilbert in his treatment of syzygies, which could be regarded as the origins of homological algebra \cite{gelfand2013methods}. More specifically, the Batalin--Vilkovisky formalism  provides a resolution (loosely speaking) of the space (of gauge-equivalence classes of) on-shell field configurations using the Koszul--Tate resolution, see e.g.~\cite{Henneaux:1992ig}. We are being a little cavalier with the term `resolution' here as the Batalin--Vilkovisky operator may have nontrivial cohomology away from degree zero (unlike the Koszul--Tate resolution itself). Really, we ought speak of arbitrary complexes up to quasi-isomorphism using the theory of derived categories. A standard example of this phenomenon is  provided by the Tor and Ext functors in homological algebra.} to that of \(Q\)-exactness.
That is, we do not care if \(\mathrm dJ\) is nonzero so long as it is \(Q\)-exact, that is, if there exists a \((d-p)\)-form \(J^{(1)}\) such that
\begin{equation}\label{eq:descent_equation_beginning}
    \mathrm dJ = QJ^{(1)}.
\end{equation}
Note that \(J^{(1)}\) has ghost number \(-1\) if \(J\) carried ghost number zero. 

This notion of a `$Q$-exact resolution' is a core principle in the Batalin--Vilkovisky formalism,  but let us motivate it specifically in the context of $p$-form symmetries. Suppose that the space of   observables in your theory are defined by the cohomology of some  operator, $Q$, that acts as a graded derivation\footnote{If $Q$ is not a differential, but rather squares to some other symmetry operator, then one has to pass to equivariant  cohomology.} (in the standard applications of the Batalin--Vilkovisky formalism these correspond to the gauge-choice independent observables). If all correlation functions of  observables  are annihilated by $Q$, then $QJ=0$ and  $\mathrm dJ = QJ^{(1)}$  are sufficient for  \eqref{eq:psym} to be topological in the sense that it is insensitive to continuous deformations of $\Sigma_{d-p-1}$. To illustrate this,  consider  the simple case of oriented  $\Sigma_{d-p-1}$ and $\Sigma'_{d-p-1}$ with oriented cobordism $\Sigma_{d-p}$ connecting them,  $(-\Sigma_{d-p-1})\cup \Sigma'_{d-p-1}=\partial \Sigma_{d-p}$. Then, as an operator equation, we have 
    \begin{equation}
\begin{aligned}
U\left(\Sigma_{d-p-1}^{\prime}\right) & =U\left(\Sigma_{d-p-1}\right) \times \exp \left(\mathrm i \alpha \int_{\partial \Sigma_{d-p}}J\right) \\
& =U\left(\Sigma_{d-p-1}\right) \times \exp \left(\mathrm i \alpha \int_{\Sigma_{d-p}} \mathrm d J\right) \\
& =U\left(\Sigma_{d-p-1}\right) \times \exp \left(\mathrm i \alpha \int_{\Sigma_{d-p}} QJ^{(1)}\right)\\
& =U\left(\Sigma_{d-p-1}\right),
\end{aligned}
\end{equation}
    so that the higher-form symmetry operator is invariant under continuous deformations as required.
    
The equation \eqref{eq:descent_equation_beginning} naturally appears in the theory of topological twists of supersymmetric field theories \cite{Labastida:2005zz} and in the theory of anomalies \cite{Bertlmann:1996xk,Harvey:2005it,Bilal:2008qx}, where one iterates  beginning with \(J^{(0)}=J\) so that
\begin{equation}\label{eq:descent_equation}
    \mathrm dJ^{(i)}=QJ^{(i+1)}.
\end{equation}
This defines a series of operators \(J^{(1)}\), \(J^{(2)}\), and so on, of increasing form degree and decreasing ghost number. Eventually the series stops when \(\mathrm dJ^{(i)}=0\) for some \(i\) (if only because form degree is bounded by the dimension of spacetime).
Now, well known arguments from the Witten-type TQFTs literature \cite{Labastida:2005zz} imply that, if \(J^{(0)}\) is a topological observable, then its descendants \(J^{(i)}\) are all topological observables; furthermore \eqref{eq:descent_equation} ensures that they are all closed up to \(Q\)-exact terms. That is, the \(J^{(i)}\) may also be regarded as generating higher-form symmetries: we may summarise this as \emph{higher-form symmetries beget lower-form symmetries}. The descent equation \eqref{eq:descent_equation} applies equally well in the case of lattice theories, where we may work with a coefficient ring such as \(\mathbb Z_q\coloneqq \mathbb Z/q\mathbb Z\).

An immediate objection to the above perspective is that such conserved `currents' have nonzero ghost number.
We reply as follows. First, in the case where the theory in question is obtained by twisting (topological, holomorphic or otherwise), the twisting procedure entails a reassignment of the ghost number since the Becchi--Rouet--Stora--Tyutin (BRST) operator of the twisted theory (with ghost number \(+1\)) comes from an ordinary supersymmetry generator (of ghost number \(0\)) of the original theory. Thus, operators with nonzero ghost number in the twisted theory may come from perfectly legitimate operators with ghost number zero in the untwisted theory. This situation is especially typical in cohomological topological field theories \cite{Labastida:2005zz} where such observables with nonzero ghost numbers wrap various cycles in spacetime and provide richer invariants of the smooth topology of spacetime; indeed, such observables derive from the descent equation \eqref{eq:descent_equation}.
Even if the theory in question did not arise from a twisting procedure,
theories often naturally have nontrivial Batalin--Vilkovisky cohomology in nonzero degrees \cite[§11.1.2]{Henneaux:1992ig},\footnote{
    Although a Koszul--Tate resolution of the on-shell locus of the gauge field by itself is guaranteed to not have cohomology in nonzero degrees \cite[§9.1.2,~Thm.~9.1]{Henneaux:1992ig}, this is sometimes awkward to work with.
}
and such ghostly cohomology plays important roles in the descent-equation approach to anomalies (see e.g.\ \cite{Bertlmann:1996xk,Harvey:2005it,Bilal:2008qx}), in the computation of dualities of the nonperturbative partition function, \cite{Siegel:1980jj, Duff:1980qv,Schwarz:1984wk,Witten:1995gf,Olive:2000yy, Donnelly:2016mlc,Borsten:2021pte,Borsten:2025phf}, in the kinematic algebra of colour--kinematics duality \cite{Ben-Shahar:2021zww,Borsten:2022vtg,Borsten:2023reb}, and in (higher) Chern--Simons theories \cite{Jurco:2018sby}.
Furthermore, from the formal perspective of the Batalin--Vilkovisky formalism and differential graded geometry, it is unnatural to privilege ghost number zero.

As a result, the sequence of \(p\)-form symmetry groups \(G^{[0]},G^{[1]},\dotsc\) enlarges into a two-dimensional array of symmetry groups \(G^{[p,q]}\) that describe \(p\)-form symmetries with ghost number \(q\).

\subsubsection{Wilson  and 't~Hooft lines}
In the above, we have discussed symmetries in terms of their associated Noether currents rather than operators inserted along submanifolds.
In general, given a \((d-p-1)\)-form current \(J\) valued in an Abelian group,\footnote{For non-Abelian groups, one must take path-ordered integrals or, more generally, Chen forms. See e.g.\ \cite{Kim:2019owc}.} then the corresponding operator for the conserved quantity \(Q_\Sigma\) on a \((d-p-1)\)-dimensional submanifold \(\Sigma\subset M\) of spacetime is
\begin{equation}
    Q_\Sigma\coloneqq \int_\Sigma J.
\end{equation}
When \(J\) has ghost number \(q\), in order for the argument of the exponential to have zero degree, we must take the parameter \(\alpha\) to have ghost number \(-q\). That is, the \(p\)-form symmetry group \(G^{[p,q]}\) with ghost number \(q\) is naturally a graded Lie group in degree \(q\) (such that the corresponding coordinate functions have degree \(-q\)).

If the action contains a term
\begin{equation}
    S = \int \dotsb + J\wedge\mathrm dW
\end{equation}
for some \(p\)-form operator \(W\), then given a closed \(p\)-dimensional submanifold \(\Sigma\), one can construct the Wilson operator \(\int_\Sigma W\), on which the symmetry \(J\) acts. This statement continues to hold when \(J\) and hence \(W\) have nonzero ghost numbers following the usual arguments, and one can discuss screening of charges in the same language as well.

Similarly, the usual arguments for higher-form symmetries persisting under renormalisation group flow applies to ghostly symmetries as well if one keeps track of the `renormalisation' of fields with nonzero ghost number.

\subsubsection[\((-1)\)-form symmetry]{\(\boldsymbol{(-1)}\)-form symmetry}
In a \(d\)-dimensional boundaryless spacetime \(M\),
\emph{any} top-degree form \(\alpha\in\mathcal A^{d,q}\) (for arbitrary ghost number \(q\)) is trivially the Noether current for a \((-1)\)-form symmetry.
In this case, no Wilson or 't~Hooft operator can transform under it for degree reasons since no \((-1)\)-dimensional submanifolds exist.
However, such \((-1)\)-form symmetries arise naturally as descendants of higher-form symmetries. See \cite{Vandermeulen:2022edk,Heckman:2024oot, Santilli:2024dyz, Najjar:2024vmm} for previous work  on (ghost number zero) \((-1)\)-form symmetries.

\subsection{Cohomological formalisation}
Let us discuss how \eqref{eq:descent_equation} arises from the Batalin--Vilkovisky cohomology.
We first consider the space of differential-form operators \(\alpha\in\mathcal A\) that are (a) closed up to $Q$-exact terms and (b) gauge-invariant in the Batalin--Vilkovisky formalism. This corresponds to the cohomology
\begin{equation}
    \operatorname H^{\bullet}_{\mathrm d}(\operatorname H_Q^{\bullet}(\mathcal A)),
\end{equation}
which is the second page of the horizontal spectral sequence approximating the total cohomology of the double complex $\mathcal A^{\bullet, \bullet}$.

We next relax this requirement to those operators \(\alpha\in\mathcal A\) that are (a) closed up to $Q$-exact terms and (b\('\)) \emph{not} necessarily physical but whose integral \(\int_\Sigma\alpha\) is physical as long as \(\Sigma\) does not have a boundary. We will find that the descent equation \eqref{eq:descent_equation} precisely captures the latter requirement of (a)+(b\('\)). 

Solving the descent equations imposed on all operators for all  total degrees $n=p+q$, is  precisely equivalent to computing the cohomology of the total cochain complex
$\operatorname H^\bullet_{\operatorname D}(\operatorname{Tot}^{\bullet}(\mathcal A))$, where $\operatorname{Tot}^{\bullet}(\mathcal A)$ is the total complex of the double complex $\mathcal A^{p, q}$,
\begin{equation}
\operatorname{Tot}^{n}(\mathcal A) \coloneq \bigoplus_{n=p+q} \mathcal A^{p, q},
\end{equation}
with total differential $\on D = \operatorname d+(-1)^\Upsilon Q$ (where $(-1)^\Upsilon\alpha = (-1)^q\alpha$ for $\alpha\in \mathcal{A}^{p,q}$ and $\operatorname D^2=0$ follows from $\on d Q=Q\on d$). For a TQFT (or a topological subsector of a generic QFT), the descent equations hold identically for all  $n=p+q$ so that the topological operators are given precisely by $\on H^\bullet_{\on D}(\on{Tot}^{\bullet}(\mathcal A))$. 

For a generic QFT, on the other hand, it might be that  the descent equations hold identically for some  but not all operators of total degree $n$.  Imposing the descent equations for those cases where they do not hold identically will reduce the space of operators to a topological subsector. However, for constructing higher-form symmetries it is sufficient to only consider those descent equations that hold identically; when they fail to do so, the sequences of symmetries simply terminates. This leads us to the `current cohomology', denoted $\on{Curr}(\mathcal A)$ and given in \eqref{eq:curr}, as the space characterising the ghostly higher-form symmetries of a QFT (without any restrictions on the operators).

\subsubsection{Cohomology of local operators}
According to the Batalin--Vilkovisky formalism, physical observables belong to the \(Q\)-cohomology; that is, the space of differential-form-valued physical local observables is the cohomology
\begin{equation}
    \operatorname H^{\bullet}_Q(\mathcal A).
\end{equation}
This includes, however, \emph{all} differential-form-valued observables, not just the \(\mathrm d\)-closed ones; we would also like to quotient out by \(\mathrm d\)-exact observables since they do not yield nonzero conserved quantities, so that one obtains
\begin{equation}
    \operatorname H^{\bullet}_{\mathrm d}(\operatorname H^{\bullet}_Q(\mathcal A)),
\end{equation}
which corresponds to the second page of the horizontal spectral sequence of the double complex $\mathcal A^{\bullet, \bullet}$.
This space may be more explicitly represented as
\begin{itemize}
\item the space of elements \(\alpha\in\mathcal A\) that are \(Q\)-closed and \(\mathrm d\)-closed up to a \(Q\)-exact element, i.e.
\begin{align}\label{eq:stronger_cond}
Q\alpha &=0, &
\exists\beta\in\mathcal A\colon \quad\mathrm d\alpha=Q\beta,
\end{align}
\item modulo linear combinations of those elements that are (\(\mathrm d\)-exact and \(Q\)-closed) \emph{or} \(Q\)-exact, i.e. those elements \(\alpha\in \mathcal A\) such that
\begin{equation}
    \exists\beta,\gamma\in\mathcal A\colon\quad\left(\alpha=\mathrm d\beta+Q\gamma\text{ and }Q\beta=0\right).
\end{equation}
\end{itemize}
Notice that \eqref{eq:stronger_cond} contains the descent equation \eqref{eq:descent_equation} but is stronger than it.

Note, the cohomology  $\operatorname H^{\bullet}_{\mathrm d}(\operatorname H^{\bullet}_Q(\mathcal A))$ admits a standard presentation in terms of pairs $(\alpha, \beta)\in \mathcal A^{p,q}\times  \mathcal A^{p+1,q-1}$ such that
\begin{equation}
Q\alpha=0, \qquad \mathrm d\alpha=Q\beta, 
\end{equation}
modulo pairs of the form
\begin{equation}
(0, \beta), \qquad (Q \gamma, \mathrm d \gamma), \qquad (\mathrm d \omega, 0), 
\end{equation}
where $\gamma \in \mathcal A^{p,q-1}$ and $\omega\in \mathcal A^{p-1,q}$ is $Q$-closed $Q\omega =0$. The first condition says all currents are `gauge-invariant' (really, independent of gauge fixing) and $\mathrm d$-closed up to  `gauge' transformations. The first relation says that all pairs of currents $(J^{(i)}, J^{(i+1)})$ and $(J^{(i)}, J'^{(i+1)})$ are equivalent, the second says that trivial solutions to the descent equation $\mathrm d Q \omega= Q\mathrm d \omega$ are trivial  in cohomology and all $\mathrm d$-exact currents $J$ are trivial.

\subsubsection{Cohomology of integrals of local operators}\label{ssec:int_coho}
Instead of requiring that \(\alpha\in\mathcal A\) be itself physical (i.e.~belong to the \(Q\)-cohomology), we may simply require that \(\int_\Sigma\alpha\) belong to the \(Q\)-cohomology whenever \(\Sigma\) is a submanifold without boundary\footnote{When $\partial \Sigma\not= \varnothing$, then one may be able to include a boundary operator  to cancel the resulting $i^*\beta$ on $\partial \Sigma$ (this is effectively the BV-BFV formalism, reviewed in \cite{Cattaneo:2019jpn}, applied to $\Sigma$ ).}. In order to have
\begin{equation}
    Q\int_\Sigma\alpha
    =\int_\Sigma Q\alpha
    =0,
\end{equation}
it suffices by Stokes' theorem to require that \(Q\alpha\) is \(\mathrm d\)-exact, i.e.
\begin{equation}
    \exists\beta\colon\quad Q\alpha=\mathrm d\beta.
\end{equation}
This is weaker than \eqref{eq:stronger_cond} and is nothing more than the descent equation \eqref{eq:descent_equation} but applied in the other direction (i.e.~requiring \(Q\alpha\)-closed up to \(\mathrm d\)-exact terms rather than \(\mathrm d\alpha\)-closed up to  \(Q\)-exact terms).
We must, however, still quotient out by \(Q\)-exact such operators (which are pure gauge), as well as by \(\mathrm d\)-exact such operators, since if \(\alpha\) is \(\mathrm d\)-exact and \(\Sigma\) is without boundary then \(\int_\Sigma\alpha=0\).

Thus, we see that the space of differential-form operators \(\alpha\) that are closed up to $Q$-exact terms and whose integral \(\int_\Sigma\alpha\) is physical is given by the cohomology
\begin{equation}\label{eq:curr}
    \operatorname{Curr}(\mathcal A)\coloneqq
    \operatorname H^{\bullet}_{\mathrm d}\mleft(\mathcal A/\operatorname{im}_{\mathcal A}(Q)\mright)
    \cap
    \operatorname H^{\bullet}_Q\mleft(\mathcal A/\operatorname{im}_{\mathcal A}(\mathrm d)\mright)
    \subset
  \frac{\mathcal A}{\left(\operatorname{im}_{\mathcal A}(Q)+\operatorname{im}_{\mathcal A}(\mathrm d)\right)},
\end{equation}
where both \(\operatorname H^{\bullet}_{\mathrm d}\mleft(\mathcal A/\operatorname{im}_{\mathcal A}(Q)\mright)\) and \(\operatorname H^{\bullet}_Q\mleft(\mathcal A/\operatorname{im}_{\mathcal A}(\mathrm d)\mright)\) are regarded as subsets of \(\mathcal A/\left(\operatorname{im}_{\mathcal A}(Q)+\operatorname{im}_{\mathcal A}(\mathrm d)\right)\).
This space may be more explicitly represented as
\begin{itemize}
\item the space of elements \(\alpha\in\mathcal A\) that are \(\mathrm d\)-closed up to a \(Q\)-exact element, i.e.
\begin{align}
\exists\beta\in\mathcal A\colon \quad\mathrm d\alpha=Q\beta,
\end{align}
as well as \(Q\)-closed up to a \(\mathrm d\)-exact element, i.e.
\begin{align}
\exists\gamma\in\mathcal A\colon \quad Q\alpha=\mathrm d \gamma,
\end{align}
\item modulo linear combinations of those elements that are either \(\mathrm d\)-exact or \(Q\)-exact, i.e. those elements \(\alpha\in \mathcal A\) such that
\begin{equation}\label{eq:gauge_trivial_current}
    \exists\beta,\gamma\in\mathcal A\colon\quad \alpha=\mathrm d\beta+Q\gamma.
\end{equation}
\end{itemize}

If we iteratively solve the descent equation by starting with an element \(\alpha\in\mathcal A^{p,q}\) to try to obtain \(\alpha^{(1)}\in\mathcal A^{p+1,q-1},\alpha^{(2)}\in\mathcal A^{p+2,q-2},\dotsc\) as well as \(\alpha^{(-1)}\in\mathcal A^{p-1,q+1},\alpha^{(-2)}\in\mathcal A^{p-2,q+2},\dotsc\), then we can phrase the above criterion as: symmetries are those where we can go up a rung as well as down a rung (i.e.\ both \(\alpha^{(1)}\) and \(\alpha^{(-1)}\) exist; they may possibly be zero).

It is often the case that the descent chain extends in both directions (e.g.\ by ending with \(\alpha^{(k)}=\alpha^{(-k)}=0\) for some sufficiently large \(k\)); this is the case in the observables of Donaldson--Witten theory \cite{Labastida:1997pb}. However, this need not be the case. If the descent equation stops holding identically (i.e.\ one either cannot go up a rung or cannot go down a rung without restricting the space of observables), then at the end of the chain one has an operator that does \emph{not} correspond to a symmetry (an example is the operator \(A\) in Maxwell theory in \cref{ssec:maxwell}), whereas away from the endpoints one indeed has a symmetry (\cref{fig:symmetry_chain}).
If the theory is topological, however, since \emph{every} physical operator is topological, the descent chain cannot end in this way --- if an operator is \(Q\)-closed up to \(\mathrm d\)-exact terms (i.e.\ is physical), then it must be \(\mathrm d\)-closed up to \(Q\)-exact terms (i.e.\ must be topological).
Thus, the descent chain always extends indefinitely in downward (eventually hitting zero due to form degree reasons); indeed, the descent chains never stopping is equivalent to saying that, insofar as the differential-form operators are concerned, the theory is topological. For a general QFT, if the descent equation fails to hold identically at some point, one can still solve it, but at the expense of restricting the theory to a topological subsector.
\begin{figure}
\Large
\begin{equation*}
    \xcancel{\alpha^{(m)}}\leftrightarrow\alpha^{(m+1)}\dotsb\leftrightarrow\alpha^{(i)}\leftrightarrow \dotsb\leftrightarrow\alpha^{(n-1)}\leftrightarrow\xcancel{\alpha^{(n)}}
\end{equation*}
\caption{In a descent chain, the endpoints \(\alpha^{(m)}\) and \(\alpha^{(n)}\) are not symmetries if the chain cannot continue, but the middle terms \(\alpha^{(m+1)},\dotsc,\alpha^{(n-1)}\) are always symmetries.}\label{fig:symmetry_chain}
\end{figure}

From this point of view, the moduli space of solutions to the descent equations (for each fixed total degree $n=p+q$) and the associated   (ghostly) higher-form symmetries  correspond precisely to the cohomology of the  total complex $\operatorname{Tot}^{\bullet}(\mathcal A)$. More explicitly,  for a putative family line of (ghostly) currents of total degree $n$ one considers $\alpha\in \operatorname{Tot}^{n}( \mathcal A)$, i.e.
\begin{equation}
\alpha = \sum_{k=0}^{d} \alpha_{k,n-k},
\end{equation}
where $\alpha_{p,q}\in \mathcal A^{p,q}$.  That is, $\alpha$ packages up all currents of total degree $n$ to be related by  the descent equations. The total complex cohomology consists closed $n$-forms, $\on D\alpha=0$, modulo exact $n$-forms, $\on D\beta$. Unpacking  
$
\on D \alpha=0 
$
we have $n+1$ independent (for  degree reasons) conditions
\begin{equation}
\on d \alpha_{k,n-k} = (-)^{k}Q \alpha_{k+1,n-k-1}, \quad k=-1, 0,\ldots n,
\end{equation}
where it is understood that $\alpha_{-1,q}=0$. 
Upon setting $n=p+q$ and letting $k=p$ or $k=p-1$ these are 
\begin{equation}
\on d \alpha_{p,q}  = (-)^{p} Q \alpha_{p+1,q-1},
\qquad
\on d \alpha_{p-1,q+1} = (-)^{p-1} Q \alpha_{p,q},
\end{equation}
so that we recover precisely the conditions for the charge $Q_{\Sigma}^{p,q}=\int_\Sigma \alpha_{p,q}, \forall p,q$ to be a gauge-invariant symmetry generator. We still require $Q_{\Sigma}^{p,q}$ to be a non-trivial, however. In particular, if $\alpha_{p,q} =\on d\beta_{p-1,q}$, then the charge vanishes identically, $Q_{\Sigma}^{p,q}=0$,  and if $\alpha_{p,q} = Q\beta_{p,q-1}$, then  $Q_{\Sigma}^{p,q}=Q\int_{\Sigma} \beta_{p,q-1}$ has trivial  action (on correlators of $Q$-closed observables). 
Unpacking a trivial ($\on D$-exact) $n$-form,  $\alpha=\on D \beta$, we have  
\begin{equation}
\alpha_{0,n} + \alpha_{1,n-1} + \alpha_{2,n-2} + \cdots  =  Q \beta_{0,n-1} + \on d\beta_{0,n-1} + Q\beta_{1,n-2} +\on d\beta_{1,n-2}+\cdots
\end{equation}
At first sight, it seems that, for example, $\beta_{0,n-1}$ is used twice, once in the $Q$-exactness of $\alpha_{0,n}$ and once in the $\on d$-exactness of $\alpha_{1,n-1}$. However, recall 
\begin{equation}
\on d \alpha_{0,n} =Q \alpha_{1,n-1}.
\end{equation}
so that $\alpha_{0,n} = Q \beta_{0,n-1} $ implies $Q \alpha_{1,n-1} = Q \on d \beta_{0,n-1}$, which in turn implies $\alpha_{1,n-1} =  \on d \beta_{0,n-1} + \gamma_{1,n-1}$ where $Q\gamma_{1,n-1}=0$. Now, $\on d \alpha_{1,n-1}  = \on d \gamma_{1,n-1} = Q \alpha_{2,n-2}$ and so on up the chain. So, in summary, solving the descent equations is equivalent to the cohomology of the total complex. For a TQFT, $\mathrm D\alpha =0$ holds identically for all total degree $n$. For a generic QFT, $\mathrm D\alpha =0$ may not hold identically for all $n$. In such cases, solving $\mathrm D\alpha =0$ will restrict the space of observables to a topological subsector. A simple example of this is given by Maxwell theory, as described in \cref{ssec:maxwell}, where for $n=d-2$ and  $2$, $\mathrm D\alpha =0$ holds identically, while for $n=1$ it restricts theory to the topological sector of flat connections $F=0$. But, let us emphasise again that for the higher symmetries to be well defined one can simply neglect those terms in $\mathrm D\alpha =0$ that fail to hold identically, so that no restriction is placed on the space of operators. In the Maxwell example for total degree $n=1$, this amounts to killing the higher-form symmetry deriving from  flat  connections, while  retaining a new ghostly zero-form symmetry deriving from the ghost current $c$ associated to the BRST gauge transformation $\delta A = \mathrm dc$.

For Abelian theories at least, we may further understand the non-topological case in terms of the total complex. Let us assume that for total degree greater than $n$, the descent equations hold identically (if the ghost degree is bounded, this is always true for some $n$). Now consider the total form $\alpha$ of total degree less than $n$. If $\mathrm D$-closure does not hold for $\alpha$ identically, then the violation in Abelian theories are of the form
\begin{equation}
\mathrm D \alpha =\on d \beta + \star \gamma
\end{equation}
where $\gamma = \sum^{n+1}_{k=0} \gamma_{d-k, n+1 - k}$ and $\gamma_{d-k, n+1 - k}\in \mathcal A^{d-k, n+1 - k}$. Applying $\mathrm d$  we observe 
\begin{equation}
 \sum_{k=0}^{n}(-1)^{k}Q( \mathrm d \alpha_{k,n-k})=\sum^{n+1}_{k=0}  \mathrm d\star  \gamma_{d-k, n+1 - k} 
\end{equation} 
so that we obtain a new  set of ``magnetic'' descent equations that start with
\begin{equation}
Q((-1)^{k} \mathrm d \alpha_{k,n-k})  =\mathrm  d(\star  \gamma_{d-k, n+1 - k} )
\end{equation}
Note that after the first rung the descent equations are necessarily trivial since at the next step we have a $\mathrm d$-exact form $(-1)^{k} \mathrm d \alpha_{k,n-k}$. This is the case for Maxwell theory and Abelian higher gauge theory.

\paragraph{Spans between symmetry groups}
Consider the space of solutions \((\alpha,\beta)\) to the descent equation
\begin{equation}
    \mathrm d\alpha=Q\beta,
\end{equation}
where
\begin{equation}
    \operatorname{\widetilde{Curr}}(\mathcal A)
    =\left\{\vphantom\sum\alpha\in\mathcal A\middle|\alpha+\left(\operatorname{im}_{\mathcal A}(Q)+\operatorname{im}_{\mathcal A}(\mathrm d)\right)\in\operatorname{Curr}(\mathcal A)\right\}
\end{equation}
is the space of representatives of currents.
Then there are clearly forgetful maps
\begin{equation}
    \alpha\leftarrow(\alpha,\beta)\to\beta
\end{equation}
to representatives of the currents \(\alpha\) and \(\beta\).
It is convenient to quotient out by \(\mathrm d\)- and \(Q\)-exact\footnote{
    If we quotient out by \(\mathrm d\)- and \(Q\)-closed terms instead, then the maps \eqref{eq:projection_solutions} are no longer well defined. 
} terms on the left and right respectively, to obtain a smaller space of solutions
\begin{equation}
    P = \frac{
        \{(\alpha,\beta)\in\operatorname{\widetilde{Curr}}(\mathcal A)\times\operatorname{\widetilde{Curr}}(\mathcal A)|\mathrm d\alpha=Q\beta\}
    }{
        \operatorname{im}_{\mathcal A}(\mathrm d)\times 
        \operatorname{im}_{\mathcal A}(Q)
    }
\end{equation}
together with forgetful maps
\begin{equation}\label{eq:projection_solutions}
    \operatorname H^{\bullet}_{\mathrm d}(\mathcal A/\operatorname{im}_{\mathcal A}(Q))\leftarrow
    P
    \to
    \operatorname H^{\bullet}_{\mathrm d}(\mathcal A/\operatorname{im}_{\mathcal A}(Q))
\end{equation}
to the symmetry currents themselves rather than their representatives.
Restoring the grading and exponentiating, we obtain a pair of group homomorphisms amongst the symmetry groups themselves:
\begin{equation}
    G^{[p,q]}\leftarrow G^{[p,q];[p-1,q-1]}\to G^{[p-1,q-1]},
\end{equation}
where \(G^{[p,q];[p-1,q-1]}\) is obtained by exponentiating (homogeneous subspaces of) \(P\).
For brevity, we will call this pair a \emph{span} relating the symmetry groups \(G^{[p,q]}\) and \(G^{[p-1,q-1]}\). (The terminology comes from category theory \cite[§XII.7]{maclane}.)

\section{Scalar theory: particle conservation and its descendant}\label{sec:free_matter}
\begin{figure}
\begin{equation*}
    \begin{tikzcd}[row sep={\pgfkeysvalueof{/tikz/commutative diagrams/column sep/large},between origins}, /tikz/column sep=
         {\pgfkeysvalueof{/tikz/commutative diagrams/column sep/huge},between origins}]
         \text{$\hphantom{-()}1$-form}&&&&1\dlar[leftrightarrow]\\
         \text{\(\hphantom{-()}0\)-form}&&&\underset{\star\mathrm d\phi}{\operatorname U(1)}\dlar[leftrightarrow]\\
         \text{\((-1)\)-form}&&\underset{\phi^+}{\operatorname U(1)}\dlar[leftrightarrow]\\
         \text{\((-2)\)-form}&1\\
         &\text{g.n.\ \(-2\)}&\text{g.n.\ \(-1\)}&\text{g.n.\ \(0\)}&\text{g.n.\ \(1\)}
    \end{tikzcd}
\end{equation*}
\caption{Ghostly higher-form symmetries of a free scalar field theory.
The above does not exhaust the symmetries: for example, any top-degree form such as \(\star\phi\) is automatically closed due to degree reasons and defines a \((-1)\)-form symmetry.}\label{fig:free_field_symmetries}
\end{figure}
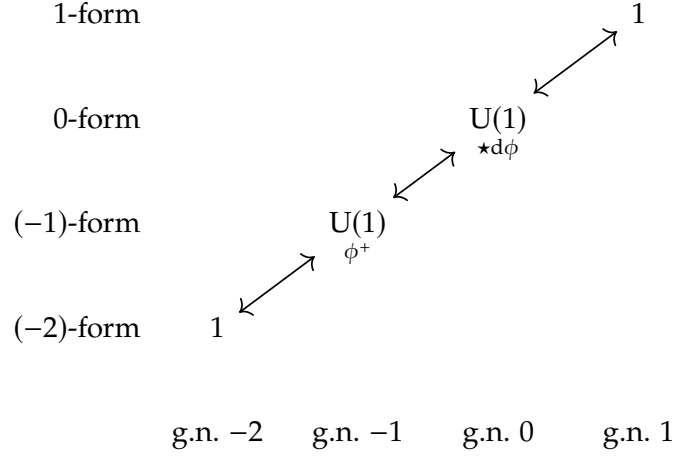

As a simple illustration of the above ideas, let us consider the theory of a free scalar field \(\phi\in\Omega^0(M)\). (To avoid an abundance of $(-1)^t$ insertions, we assume $M$ Euclidean throughout). In the Batalin--Vilkovisky formalism, one adjoins to it the corresponding antifield \(\phi^+\in\Omega^d(M)\), which we take to be a top-degree form with ghost number \(-1\). The algebra of local operators is generated by \(\phi\) and \(\phi^+\) under exterior derivatives, wedge products and Hodge stars. The Batalin--Vilkovisky differential \(Q\) may be read off from the action
\begin{equation}
    S = \int\frac12\mathrm d\phi\wedge\star\mathrm d\phi
\end{equation}
and is
\begin{align}
	Q\phi &= 0\,,
	&
	Q\phi^+ &= \mathrm{d}\star\mathrm{d}\phi\,.
\end{align}
In the usual account of symmetries, one notes that the \((d-1)\)-form \(J = \star\mathrm d\phi\) is closed up to the equations of motion:
\begin{equation}\label{eq:scalar_field_eom}
    \mathrm dJ \overset{\textsc{eom}}= 0.
\end{equation}
Thus, it is the Noether current for a zero-form symmetry, whose conserved quantity is the number of particles.
In the Batalin--Vilkovisky formalism, \eqref{eq:scalar_field_eom} is resolved into the following:
\begin{equation}
    \mathrm dJ=Q\phi^+.
\end{equation}
Thus we see that \(J=J^{(0)}\) begets its descendant \(J^{(1)}=\phi^+\), which defines a \((-1)\)-form symmetry. The chain of descendants trivialises here since
\begin{equation}
    \mathrm dJ^{(1)}=0
\end{equation}
due to degree reasons. Similarly, when going in the reverse direction, one has
\begin{equation}
    0 = QJ,
\end{equation}
so that \(J^{(-1)}=0\), with the chain trivialises in the other direction also. Thus, the family of symmetries of the free scalar field is given in \cref{fig:free_field_symmetries}.

{To place this discussion in the general cohomological context developed in \cref{ssec:int_coho}, we pass to the  first order formalism, introducing the ghost number $0$ auxiliary $(d-1)$-form operator $\chi$ and its ghost number $-1$  $1$-form antifield $\chi^+$. The action is given by
\begin{equation}
	S 
	=
	\int \chi\wedge\mathrm{d}\phi - \tfrac12 \chi\wedge\star\chi
\end{equation}
where
\begin{align}
	Q\phi &= 0\,,
	&
	Q\phi^+ &= (-)^{d-1}\mathrm d \chi\,,
	& 
	Q\chi &=0 \,,
	&
    Q\chi^+
    &=  
    -\mathrm d \phi + \star \chi\,.
\end{align}
There is  the ``electric'' total degree $(d-1)$ total form,
\begin{equation}
	\alpha^\mathrm{e}
	=
	\chi + \phi^+\,,
\end{equation}
 which identically satisfies
\be
\mathrm D \alpha^\mathrm{e} = 0
\ee
 yielding  the descent equations for currents 
\begin{align}\label{eq:scalar_electric_descent}
    0&=QJ_{\mathrm{e}}^{(0)}&
    \mathrm dJ_{\mathrm{e}}^{(0)}&=Q J_{\mathrm{e}}^{(1)},&
    \mathrm dJ_{\mathrm{e}}^{(1)}&=0&
\end{align}
for the ghostly currents 
\begin{align}\label{eq:scalar_descendants}
    J_{\mathrm{e}}^{(-1)}&=0,&
    J_{\mathrm{e}}^{(0)}&=(-)^{d-1}\chi,&
    J_{\mathrm{e}}^{(1)}&=\phi^+,&
    J_{\mathrm{e}}^{(2)}&=0.
\end{align}
This is just a rephrasing of the zero-form symmetry particle number conservation current and its ghostly $(-1)$-form symmetry descendant. There is also the total degree $0$ total form, 
\begin{equation}
	\alpha
	=
	\phi - \chi^+
\end{equation}
which yields
\begin{equation}
	\mathrm{D}\alpha = -\mathrm{d}\chi^+ + \star\chi
\end{equation}
Thus imposing $\mathrm D \alpha=0$ would imply $\chi=0$, which on-shell is $\mathrm \phi=\text{const}$. However, as described in \cref{ssec:int_coho}, this violation implies the ``magnetic'' descent equations
\begin{align}
	\mathrm{d}\mathrm{D}\alpha
	&=
	Q\big(\mathrm{d}(-)^\Upsilon\alpha\big)
	&
	&\Rightarrow
	&
	0 &= Q(\mathrm{d}\phi)\,,
	&
	\mathrm{d}(\star\chi) &= Q(\mathrm{d}\chi^+)\,.
\end{align}

It is immediate that similar considerations hold for e.g.\ a free fermion \(\psi\), so that the zero-form  conserved vector current for particle number \(J^{(0)}=\star\bar\psi\gamma_\mu\psi\,\mathrm dx^\mu\) admits a descendant \(J^{(1)}=\psi^+\) corresponding to a \((-1)\)-form symmetry. For $d$ even there is, of course, also  the zero-form  conserved axial current  \(J_{\text{axial}}^{(0)}=\star\bar\psi\gamma^{d+1}\gamma_\mu\psi\,\mathrm dx^\mu\), which comes with an axial descendant \(J_{\text{axial}}^{(1)}=\gamma^{d+1}\psi^+\). The chiral anomaly is then inherited by the descendant, 
\be
 \langle \mathrm d J_{\text{axial}}^{(0)}J^{(0)}J^{(0)} \rangle =  \langle Q J_{\text{axial}}^{(1)} J^{(0)}J^{(0)} \rangle \not=0,
 \ee
so that we may interpret the mixed 't Hooft anomaly as a failure for the descendant triangle diagrams to be $Q$-closed. }

\section{Abelian gauge theory: electric/magnetic symmetries  and their descendants}\label{sec:abelian}
\subsection{Maxwell theory}\label{ssec:maxwell}
As the simplest example of a gauge theory, we consider Maxwell theory on \(d\) spacetime dimensions.
We work with the first-order formulation of Maxwell theory since it is this form that naturally arises from the dimensional reduction of a SymTFT in the sandwich construction \cite{Pulmann:2019vrw} and the expressions are slightly simpler (however, the usual second-order formulation of Maxwell theory gives equivalent results):
\begin{equation}\label{eq:maxwell-action}
	S[A,B,c,A^+,B^+,c^+]
	\coloneqq
	\int_M B\wedge\mathrm{d}A - \tfrac12 B\wedge\star B - \mathrm{d}c\wedge A^+
\end{equation}
with differential-form fields
\begin{equation}
\begin{aligned}
    A&\in\Omega^1(M),&B&\in\Omega^{d-2}(M),&c&\in\Omega^0(M),\\
    A^+&\in\Omega^{d-1}(M),&B^+&\in\Omega^2(M),&
    c^+&\in\Omega^d(M),
\end{aligned}
\end{equation}
which are the photon, auxiliary field, ghost, and their corresponding antifields respectively; these 
carry ghost numbers \(0\), \(0\), \(1\), \(-1\), \(-1\), and \(-2\) respectively.\footnote{In general, \(A\) is only locally defined,
but for simplicity we neglect this fact.
For the first-order formulation of Yang--Mills theory this does not affect our results since \(A\) does not appear directly in the symmetry currents (and those that do appear, such as \(B\) and \(A^+\), are globally defined).
}
Upon integrating out \(B\), one recovers the standard Maxwell action \(\int\frac12\mathrm dA\wedge\star\mathrm dA\).

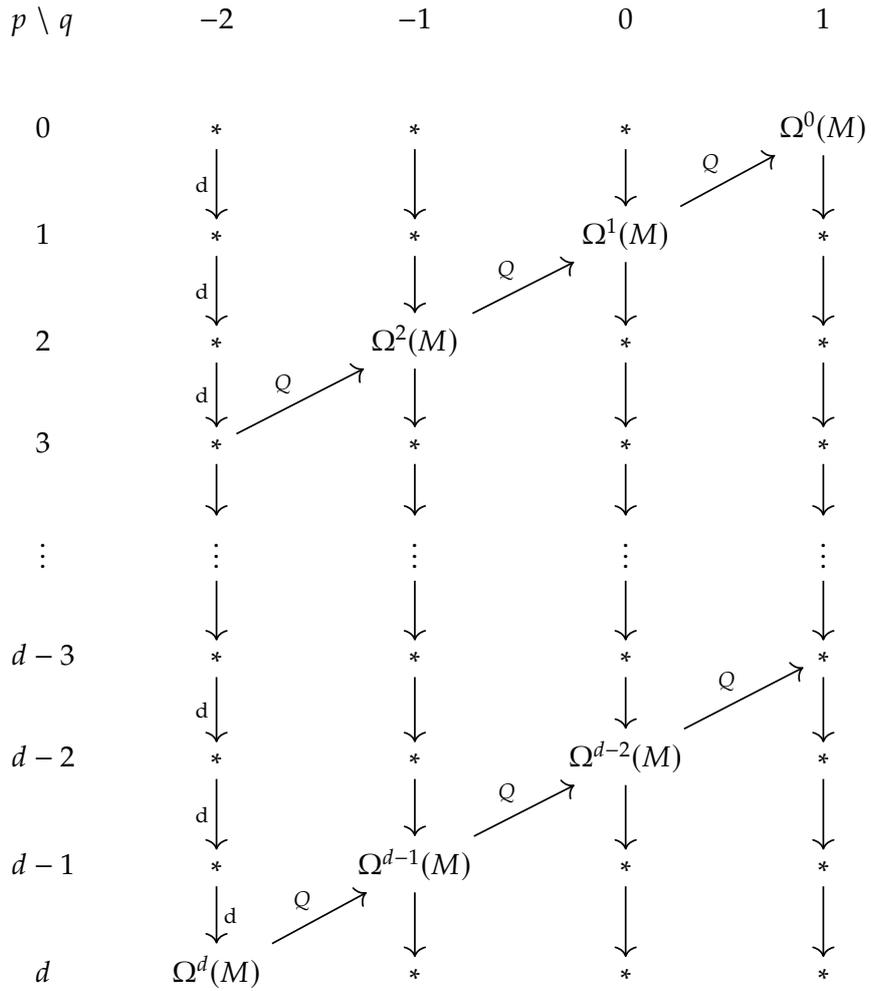
\begin{figure}
\[\begin{tikzcd}[ampersand replacement=\&]
	{p\;\backslash\;q} \& {-2} \& {-1} \& 0 \& 1 \\
	0 \& {*} \& {*} \& {*} \& {\Omega^0(M)} \\
	1 \& {*} \& {*} \& {\Omega^1(M)} \& {*} \\
	2 \& {*} \& {\Omega^{2}(M)} \& {*} \& {*} \\
	3 \& {*} \& {*} \& {*} \& {*} \\
	\vdots \& \vdots \& \vdots \& \vdots \& \vdots \\
	{d-3} \& {*} \& {*} \& {*} \& {*} \\
	{d-2} \& {*} \& {*} \& {\Omega^{d-2}(M)} \& {*} \\
	{d-1} \& {*} \& {\Omega^{d-1}(M)} \& {*} \& {*} \\
	d \& {\Omega^d(M)} \& {*} \& {*} \& {*}
	\arrow["{\mathrm d}"', from=2-2, to=3-2]
	\arrow[from=2-3, to=3-3]
	\arrow[from=2-4, to=3-4]
	\arrow[from=2-5, to=3-5]
	\arrow["{\mathrm d}"', from=3-2, to=4-2]
	\arrow[from=3-3, to=4-3]
	\arrow["Q", from=3-4, to=2-5]
	\arrow[from=3-4, to=4-4]
	\arrow[from=3-5, to=4-5]
	\arrow["{\mathrm d}"', from=4-2, to=5-2]
	\arrow["Q", from=4-3, to=3-4]
	\arrow[from=4-3, to=5-3]
	\arrow[from=4-4, to=5-4]
	\arrow[from=4-5, to=5-5]
	\arrow["Q", from=5-2, to=4-3]
	\arrow[from=5-2, to=6-2]
	\arrow[from=5-3, to=6-3]
	\arrow[from=5-4, to=6-4]
	\arrow[from=5-5, to=6-5]
	\arrow[from=6-2, to=7-2]
	\arrow[from=6-3, to=7-3]
	\arrow[from=6-4, to=7-4]
	\arrow[from=6-5, to=7-5]
	\arrow["{\mathrm d}"', from=7-2, to=8-2]
	\arrow[from=7-3, to=8-3]
	\arrow[from=7-4, to=8-4]
	\arrow[from=7-5, to=8-5]
	\arrow["{\mathrm d}"', from=8-2, to=9-2]
	\arrow[from=8-3, to=9-3]
	\arrow["Q", from=8-4, to=7-5]
	\arrow[from=8-4, to=9-4]
	\arrow[from=8-5, to=9-5]
	\arrow["{\mathrm d}", from=9-2, to=10-2]
	\arrow["Q", from=9-3, to=8-4]
	\arrow[from=9-3, to=10-3]
	\arrow[from=9-4, to=10-4]
	\arrow[from=9-5, to=10-5]
	\arrow["Q", from=10-2, to=9-3]
\end{tikzcd}
\]
\caption{The double complex $\mathcal A^{p,q}$ for Maxwell theory, where we only indicate those spaces containing $p$-form operators.}\label{fig:maxdouble}
\end{figure}

The Batalin--Vilkovisky differential \(Q\) can be read off from the action as
\begin{equation}\label{eq:Max_QBV}
\begin{aligned}
    QA\phantom{^+} &= -\mathrm dc\,,
    &
    QB\phantom{^+} &= 0\,,
    &
    Qc\phantom{^+} &= 0\,,
    \\
    QA^+&=\mathrm -\mathrm{d}B\,,
    &
    QB^+ &= -\mathrm{d}A + \star B\,,
    &
    Qc^+ &= \mathrm{d}A^+\,.
\end{aligned}
\end{equation}
It is useful to summarise this data in terms of the BV manifold regarded as a $Q$-complex:
\be\begin{tikzcd}[ampersand replacement=\&,cramped,sep=tiny]
	{\text{gh. degree}} \& {-2} \&\& {-1} \&\& 0 \&\& 1 \\
	{Q\text{-complex}} \& {\Omega^d(M)} \&\& \begin{array}{c} \begin{array}{c}\Omega^{d-1}(M)\\\oplus \\\Omega^{2}(M) \end{array} \end{array} \&\& \begin{array}{c} \begin{array}{c}\Omega^1(M)\\\oplus \\\Omega^{d-2}(M) \end{array} \end{array} \&\& {\Omega^0(M)} \\
	{\text{operator}} \& {c^+} \&\& {(A^+, B^+)} \&\& {(A,B)} \&\& c
	\arrow["Q", from=2-2, to=2-4]
	\arrow["Q", from=2-4, to=2-6]
	\arrow["Q", from=2-6, to=2-8]
\end{tikzcd}\ee
The corresponding double complex $\mathcal A^{p,q}$ of $p$-form operators is given in \cref{fig:maxdouble} and the operators $c^+, A^+, B^+, A, B, c$ carry total degree $d-2, d-2, 1, 1, d-2, 1$, respectively, while their Hodge duals carry total degree $-2, 0, d-3, d-1, 2, d+1$, respectively.

\paragraph{Conventional electric/magnetic higher-form symmetries} Maxwell theory enjoys conventional electric one-form and magnetic  $(d-3)$-form symmetries \cite{Gaiotto:2014kfa}, which each beget descendants.  The electric one-form symmetry \(\operatorname U(1)\) has Noether current  \(J_\mathrm e\coloneqq -B\) (or equivalently in the second-order formalism, ${-}{\star^{-1}} F$), which is closed on shell by the equations of motion. The magnetic \((d-3)\)-form symmetry has Noether current  \(J_\mathrm m\coloneqq \star B\) (or equivalently in the second-order formalism, $F$), which is also closed up to equations of motion. {These  provide family lines starting with conventional higher-form symmetries that beget ghostly descendants as described below.

\paragraph{Electric one-form symmetry and its descendants} The one-form electric symmetry begins with the total degree $n=(d-2)+0$ current $B$ and}  begets the following descendants:
\begin{align}\label{eq:maxwell_electric_descent}
    0 &= QJ_\mathrm e\,, 
    &
    \mathrm dJ_\mathrm e &= QA^+\,,
    &
    \mathrm d A^+ &= Q c^+\,,
    &
    \mathrm d c^+ &= 0\,.
\end{align}
Thus, the descendants of the electric Noether current are
\begin{align}\label{eq:maxwell_descendants}
    J_{\mathrm e}^{(-1)}&=0,&
    J_{\mathrm e}^{(0)}&= - B,&
    J_{\mathrm e}^{(1)}&=A^+,&
    J_{\mathrm e}^{(2)}&=c^+,&
    J_{\mathrm e}^{(3)}&=0.
\end{align}
{In terms of the cohomology of the total complex $\on{Tot}^\bullet(\mathcal A)$, we start with the electric total $n=(d-2)$-form
\begin{equation*}
	\alpha^\mathrm{e}
	=
	\sum_{k=0}^{d-2}\alpha^{\mathrm{e}}_{k,d-2-k}
	=
	c^+ + (-)^{d-1}A^+ + B
\end{equation*}
Then $\mathrm D$-closure,
\begin{equation*}
	\mathrm{D}\alpha^\mathrm{e}
	=
	\mathrm{d}c^+
	+
	(-)^d\big(Qc^+ - \mathrm{d}A^+\big)
	+
	\big(QA^+ + \mathrm{d}B\big)
	+
	QB
	= 0
\end{equation*}
is just \eqref{eq:maxwell_electric_descent}; $\mathrm D \alpha^\mathrm{e}=0$ is identically  satisfied with no restriction on the operators. The total $(d-2)$-form is to be regarded modulo $\mathrm D \beta$, where $\beta$ is a total $(d-3)$-form. However, there are generically no total $(d-3)$-forms, so the tower of ghostly symmetries is given by precisely the Noether currents of \eqref{eq:maxwell_descendants}. For $d=4$ there is a  total degree-1 polyform, that provides a  putative gauge transformation, however the conclusions remain unchanged since $A$ is, in general, not globally defined. 
}

\paragraph{The magnetic $(d-3)$-form symmetry and its descendants} The $(d-3)$-form magnetic symmetry begins with current $J_\mathrm{m} \coloneqq \star B$  and  begets the following descendants:
\begin{align}\label{eq:j_m_descendant_maxwell}
	0 &= QJ_\mathrm{m}\,,
	&
	\mathrm{d}J_\mathrm{m} &= Q(\mathrm{d}B^+)\,,
	&
	\mathrm{d}(\mathrm{d}B^+) &= 0\,.
\end{align}
Thus
\begin{align}
    J_{\mathrm m}^{(-1)}&=0,&
    J_{\mathrm m}^{(0)}&=\star B,&
    J_{\mathrm m}^{(1)}&=\mathrm dB^+,&
    J_{\mathrm m}^{(2)}&=0.
\end{align}
However, \(J_{\mathrm m}^{(1)}\) is \(\mathrm d\)-exact and hence does not represent a true physical symmetry, according to \eqref{eq:gauge_trivial_current}.

{Following the general discussion of \cref{ssec:int_coho}, in terms  of the total complex $\on{Tot}^\bullet(\mathcal A)$,  the magnetic symmetries above  follow from the violation of $\mathrm D$-closure for the  total degree $1$ total form
\begin{equation}
	\alpha
	=
	\sum_{k=0}^{1}\beta_{k,1-k}
	=
	-c + A + B^+
\end{equation}
as discussed below. 
}

\paragraph{The ghostly $(d-1)$-form symmetry and its descendants} One can also consider the final distinct family  with total degree $1$, which includes the putative currents $B^+, A$ and $c$. However, while $c$ provides a symmetry, the next descent equation for $A$ fails to hold identically, so that the family line terminates (unless one restricts to a topological sector consisting of flat connections). 

\begin{figure}
\begin{equation*}
    \begin{tikzcd}[row sep={\pgfkeysvalueof{/tikz/commutative diagrams/row sep/large},between origins}, column sep={4.5em,between origins}]
    \text{\(\hphantom{(\;-0)}d\)-form}&&&&&&1\dlar[leftrightarrow]\\
    \text{\((d-1)\)-form}&&&&&\underset c{\operatorname U(1)}\dlar[leftrightarrow]\\
    \text{\((d-2)\)-form}&&&&\xcancel{A}&1\dlar[leftrightarrow]\\
    \text{\((d-3)\)-form}&&&&\underset{\star B}{\operatorname U(1)}\dlar[leftrightarrow]\\
    \text{\((d-2)\)-form}&&&1\\
    \qquad\quad\vdots\\
    \text{\(\hphantom{()-d}2\)-form}&&&&&1\dlar[leftrightarrow]\\
    \text{\(\hphantom{()-d}1\)-form}&&&&\underset B{\operatorname U(1)}\dlar[leftrightarrow]\\
    \text{\(\hphantom{()-d}0\)-form}&&&\underset{A^+}{\operatorname U(1)}\dlar[leftrightarrow]\\
    \text{\(\hphantom{\;d\;}(-1)\)-form}&&\underset{c^+}{\operatorname U(1)}\dlar[leftrightarrow]\\
    \text{\(\hphantom{\;d\;}(-2)\)-form}&1\\
    &\text{g.n.\ \(-3\)}&\text{g.n.\ \(-2\)}&\text{g.n.\ \(-1\)}&\text{g.n.\ \(0\)}&\text{g.n.\ \(1\)}&\text{g.n.\ \(2\)}
    \end{tikzcd}
\end{equation*}
\caption{Ghostly higher-form symmetries of pure Maxwell theory in \(d\) spacetime dimensions. The crossed-out \(\xcancel A\) does not represent a symmetry but is included to indicate that the descent chain ends there. Note that, as the theory lacks charged matter, there is no zero-form symmetry with ghost number zero corresponding to charge conservation.}\label{fig:symmetries_maxwell}
\end{figure}
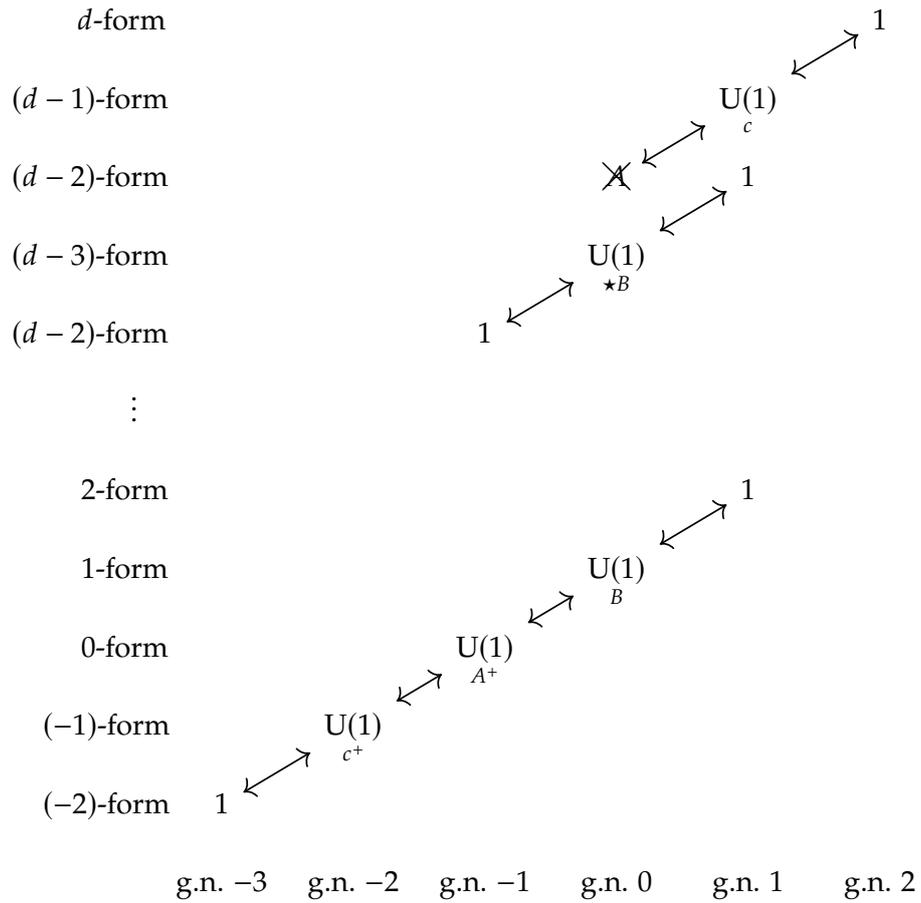

To see this, consulting \eqref{eq:Max_QBV} we conclude that \(c\) is actually a conserved current, generating a \((d-1)\)-form symmetry of ghost number \(1\), since
\begin{align}\label{eq:max_ghost_current}
    0&=Q(-c),&
    \mathrm d (-c) &= QA,&
    \operatorname{im}_{\mathcal A}(Q)&\not\ni \mathrm d A.
\end{align}
However, in this case, the chain actually stops at the next rung \(A\) since the next  descent equation does not hold identically (as  $\mathrm d A \notin \operatorname{im}_{\mathcal A}(Q)$) and would  imply \(F=\mathrm dA\) vanishes on shell. Thus, while \(c\) is a symmetry, \(A\) is not, unless one restricts flat connections and thus to a topological theory. Thus, the higher-form symmetries of pure Maxwell theory and their generators are as given in \cref{fig:symmetries_maxwell}.

{In terms of cohomology of the total complex,  for the total degree $1$ total form 
\begin{equation}
	\alpha
	=
	-c + A + B^+\,.
\end{equation}
we have
\begin{equation}
	\mathrm{D}\alpha
	=
	Qc - (\mathrm{d}c + QA) + (\mathrm{d}A + QB^+) + \mathrm{d}B^+
	=
	\mathrm{d}B^+ + \star B
\end{equation}
which includes the ghost currents descent equations \eqref{eq:max_ghost_current},
 but also yields the $(d-3)$-form magnetic symmetry 
\begin{align}
	\mathrm{d}\mathrm{D}\alpha
	&=
	Q\big(\mathrm{d}(-)^\Upsilon\alpha\big)
	&
	&\Rightarrow
	&
	0 &= QJ_\mathrm{m}\,,
	&
	\mathrm{d}J_\mathrm{m} &= Q(\mathrm{d}B^+)\,,
	&
	\mathrm{d}(\mathrm{d}B^+) &= 0\,.
\end{align}
Imposing $\mathrm D\alpha =0$ implies $B=QX$, so that theory would be restricted to the topological sector of flat connections. 
}

\subsubsection{Second-order formulation of Maxwell theory}
As an example that the ghostly higher symmetries we find are physical, let us compare the usual second-order formulation of Maxwell theory:
\begin{equation*}
	S[A,A^+,c,c^+] \coloneqq \int_M \tfrac12 \mathrm{d}A\wedge\star\mathrm{d}A - \mathrm{d}c\wedge A^+
\end{equation*}
with the Batalin--Vilkovisky differential
\begin{align}
	QA &= -\mathrm{d}c\,,
	&
	Qc &= 0\,,
	&
	QA^+ &= -\mathrm{d}\star\mathrm{d}A\,,
	&
	Qc^+ &= \mathrm{d}A^+\,.
\end{align}
Then we have as before the symmetries \(J_\mathrm e\coloneqq - \star\mathrm dA\)
and \(J_\mathrm m\coloneqq\mathrm dA\) and \(c\), whose descendants are
\begin{align}
    \mathrm dJ_\mathrm e &= QA^+\,,
    &
    \mathrm dA^+ &= Qc^+\,,
    &
    \mathrm dc^+ &= 0\,,
    &
    \mathrm dJ_\mathrm m &= 0\,,
    &
    \mathrm d(-c) &= QA\,.
\end{align}
(In the second-order formulation, whilst \(J_{\mathrm m}=\mathrm dA\) looks like it is exact, in fact it is not due to the fact that \(A\) is not globally defined, a complication that does not arise in the second-order formulation since \(B\) is globally defined, so \(J_{\mathrm m}\) still represents a physical symmetry.)
Thus one obtains the same set of ghostly higher-form symmetries.

Notice that in the second-order formulation \(J_{\mathrm m}\) has no descendants on the nose unlike the first-order formulation where \(J_{\mathrm m}\) does beget a descendant that happens to be \(\mathrm d\)-exact \eqref{eq:j_m_descendant_maxwell}.

\subsubsection{Interpretation of the ghostly zero-form and $(d-1)$-form symmetries}
The ghostly zero-form symmetry \(\operatorname U(1)^{[0,-1]}\) is an ordinary symmetry. It is nothing but the  ghost shift symmetry $c\mapsto c+\lambda$ (where $\mathrm d \lambda=0$), associated to the current \(A^+\). Indeed, from the action \eqref{eq:maxwell-action}, the equation of motion for \(c\) enforces that \(\mathrm dA^+=0\). In this regard, $c$ is the Nambu--Goldstone boson of a spontaneously broken \(\operatorname U(1)^{[0,-1]}\), just as $A$ is the Nambu--Goldstone boson of a spontaneously broken $1$-form electric symmetry \cite{Gaiotto:2014kfa}.

The corresponding conserved quantity counts the number of quanta of \(A^+\) in an extended Fock space containing antifields.
Recall that, in the BRST formalism, to show the unitarity of the scattering amplitudes, one formally considers an extended graded Fock space  that includes ghosts and antighosts \cite[§16.4]{Peskin:1995ev} with a non-positive-definite, but nondegenerate inner product, on which the BRST differential \(Q_\mathrm{BRST}\) acts; this extended `Hilbert' space\footnote{Technically, it is not a Hilbert space in the mathematical sense since the inner product is not positive-definite.} also appears in colour--kinematics duality \cite{Borsten:2020zgj,Borsten:2021hua}.
We may further extend the Fock space to also include excitations of the antifields (of arbitrary polarisation); on this extended Fock space \(\mathcal H_\mathrm{BV}\) we have an action of the Batalin--Vilkovisky differential \(Q\colon\mathcal H_\mathrm{BV}\to\mathcal H_\mathrm{BV}\), and the Batalin--Vilkovisky formalism formally produces non-unitary `dynamics' on states in this extended Fock space.
On \(\mathcal H_\mathrm{BV}\), then,
the zero-form symmetry \(\operatorname U(1)^{[0,-1]}\) corresponds to the conservation of
the number of timelike-polarised quanta of \(A^+\).

A similar analysis applies to the ghostly $(d-1)$-form symmetry \(\operatorname U(1)^{[d-1,1]}\), which is the antifield  shift symmetry $A^+\mapsto A^++\Lambda$, where $\mathrm d\Lambda=0$. In this case the current is $c$, and the conserved charge counts the number of $c$ ghost quanta  in the BRST-extended Fock space. 

\subsubsection{Wilson operators}\label{ssec:wilson_ops}
It is well known that the electric one-form symmetry with Noether current \(J_{\mathrm e}\coloneqq B\) acts on the charge $q$ Wilson line operator \(\exp\mathrm iq\int_{\Sigma_1} A\) inserted along a one-dimensional submanifold \(\Sigma_1\). This can be directly read off from the action \eqref{eq:maxwell-action}, which contains a term \(B\wedge\mathrm dA\). Similarly, we see that the action \eqref{eq:maxwell-action} contains an \(A^+\wedge\mathrm dc\) term, which shows that the ghostly zero-form symmetry with Noether current \(J_{\mathrm e}^{(1)}=A^+\) acts on the `Wilson point' \(\exp\mathrm iq\int_{\Sigma_0} c\), where \(\Sigma_0\) is a zero-dimensional submanifold (i.e.\ a point in \(M\)) -- that is, an insertion of the local operator \(c\); and dually, the ghostly \((d-1)\)-form symmetry with Noether current \(c\) acts on the Wilson surface \(\int_{\Sigma_{d-1}}A^+\) along a \((d-1)\)-dimensional submanifold \(\Sigma_{d-1}\).

The \((-1)\)-form symmetry \(J_{\mathrm e}^{(2)}=c^+\) does not act on any Wilson operator for degree reasons, since a \((-1)\)-dimensional submanifold does not exist.

For the magnetic symmetry \(J_{\mathrm m}\coloneqq\star B\), the action \eqref{eq:maxwell-action} does not contain any term of the form \(\star B\wedge\mathrm d\tilde A\) directly. However, the equations of motion force \(B\) to be closed, so that one can (at least locally) define \(B=\mathrm d\tilde A\), where \(\tilde A\) is the dual potential; in that case, the term \(\star B\wedge B=\star B\wedge\mathrm d\tilde A\) means that the Wilson hypersurface \(\exp\mathrm i\int_\Sigma \tilde A\) inserted along a \((d-3)\)-dimensional submanifold \(\Sigma\subset M\) transforms nontrivially under \(J_{\mathrm m}\), and it is clear that this is in fact a 't~Hooft operator for the original field \(A\).
Its descendant \(J_{\mathrm m}^{(1)}=\mathrm dB^+\) does not appear in \eqref{eq:maxwell-action} since \(B\) does not transform under a gauge symmetry and does not act on any Wilson or 't~Hooft operators; this comports with the fact that it is \(\mathrm d\)-exact and hence does not represent a physical symmetry.

\subsubsection{Electromagnetic duality and  ghostly symmetries}
Note that, when \(d=4\), the above ghostly symmetries are \emph{not} invariant under electromagnetic duality, which exchanges \(J_{\mathrm e}\) (which begets two descendants) and \(J_{\mathrm m}\) (which does not beget a physical symmetry). This is natural insofar as electromagnetic duality admits an interpretation in terms of sandwich SymTFTs \cite{Pulmann:2019vrw}, which in turn corresponds to gauging of certain higher-form symmetries. In this case, we see that the higher-form symmetry that is being gauged away/appearing corresponds to the ghostly \(0\)-form and \((-1)\)-form symmetries.

In general, Maxwell theory (and Abelian higher gauge theories) suffer a duality anomaly \cite{Donnelly:2016mlc} in an even number of spacetime dimensions that is sensitive to the presence (or absence) of ghosts,
so the non-invariance of higher ghost symmetries is consistent with this picture.
(For interactions between higher form symmetry and electromagnetic duality, see \cite{Hull:2024uwz}.)

\subsection{Abelian higher gauge theory}\label{ssec:higher_abelian_gauge_theory}
The previous discussion for Maxwell theory generalises readily to Abelian \(p\)-form gauge theory for arbitrary \(p\) (which subsumes the scalar and Maxwell examples).
On an oriented \(d\)-dimensional spacetime \(M\), let \(C\) be a \(\operatorname U(1)\)-valued \(p\)-form potential, and let \(B\) be a \((d-p-1)\)-form auxiliary field.
The action functional for \(p\)-form electrodynamics can the be written in a first-order fashion as
\begin{equation}
	S = \int B\wedge\mathrm{d}C - \tfrac12 B\wedge\star B\,.
\end{equation}
Integrating out \(B\) reproduces the standard action \(\int\frac12\mathrm dC\wedge\star\mathrm dC\).
The first-order form has the advantage that it naturally arises from dimensional reduction from a SymTFT in the sandwich construction \cite{Pulmann:2019vrw}.

This theory has an electric \(p\)-form electric symmetry with Noether current \(J_{\mathrm e}\coloneqq (-)^{\Upsilon(d-\Upsilon)} B\) as well as a magnetic \((d-p-2)\)-form symmetry with Noether current \(J_{\mathrm m}\coloneqq\star B\).
In the Batalin--Vilkovisky formalism, we add a tower of higher-order ghosts \(\Lambda_1,\dotsc,\Lambda_p\), so that the full Batalin--Vilkovisky action becomes
\begin{equation}\label{eq:p-form-BV-action}
	S
	=
	\int -\tfrac12\Lambda_{-1}^+\wedge\star\Lambda_{-1}^+
	- \mathrm d\Lambda_0 \wedge \Lambda_{-1}^+
    - \mathrm d\Lambda_1\wedge\Lambda_0^+
    - \dotsb
    - \mathrm d\Lambda_p\wedge\Lambda_{p-1}^+\,,
\end{equation}
where \(\Lambda_i\) is a differential \((p-i)\)-form with ghost number \(i\), and \(\Lambda_i^+\) is the corresponding antifield, a differential \((d-p+i)\)-form with ghost number \(-i-1\), and we identify
\begin{equation}
\begin{aligned}
	C\phantom{^+} &= -\Lambda_0\,,
	&
	B\phantom{^+} &= (-)^{(p+1)(d-p-1)}\Lambda^+_{-1}
	&
	&\Leftrightarrow
	&
	J_\mathrm{e} &= \Lambda_{-1}^+\,,
	&
	J_\mathrm{m} &= \star^{-1}\Lambda_{-1}^+\,,
	\\
	C^+ &= -\Lambda_0^+\,,
	&
	B^+ &= (-)^{(p+1)(d-p-1)}\Lambda_{-1}\,.
\end{aligned}
\end{equation}
The Batalin--Vilkovisky differential can be then read off as
\begin{equation}\label{eq:abelian_higher_gauge_bv_differential}
\begin{aligned}
	Q\Lambda_{-1} &= -\mathrm{d}\Lambda_0 - \star^{-1}\Lambda_{-1}^+\,,
	&
	Q\Lambda^+_{-1} &= 0\,,
	\\
	Q\Lambda_i &= -\mathrm{d}\Lambda_{i+1}\,,
	&
	Q\Lambda^+_i &= (-)^{p+1}\mathrm{d}\Lambda^+_{i-1}\,,
	&
	&(0 \leq i < p)
	\\
	Q\Lambda_p &= 0\,,
	&
	Q\Lambda^+_p &= (-)^{p+1}\mathrm{d}\Lambda^+_{p-1}\,,
\end{aligned}
\end{equation}
where \(\star^{-1} = (-)^{\Upsilon(d-\Upsilon)}\star\), as we're working in Euclidean signature. For the electric symmetry \(J_{\mathrm e}\), the descent equation produces the following descendants:
\begin{equation}\label{eq:pform_elec_descent}
\begin{aligned}
	0 &= QJ_\mathrm{e}
	\\
	\mathrm{d}J_\mathrm{e} &= (-)^{p+1}Q\Lambda^+_0 
	\\
	\mathrm{d}\Lambda^+_0 &= (-)^{p+1}Q\Lambda^+_1 
	\\
	&\vdots
	\\
	\mathrm{d}\Lambda_{p-1}^+ &= (-)^{p+1}Q\Lambda^+_p 
	\\
	\mathrm{d}\Lambda_p^+ &= 0
\end{aligned}
\end{equation}
The magnetic symmetry \(J_{\mathrm m}\), on the other hand, only begets one descendant:
\begin{align}
	0 &= QJ_\mathrm{m}\,,
	& 
	\mathrm{d}J_\mathrm{m} &= Q(-\mathrm{d}\Lambda_{-1})\,,
	&
	\mathrm{d}(-\mathrm{d}\Lambda_{-1}) &= 0\,.
\end{align}
The descendant \(J^{(1)}_\mathrm m= -\mathrm d\Lambda_{-1}\) is \(\mathrm d\)-exact and hence according to \eqref{eq:gauge_trivial_current} does not represent a physical symmetry.
In addition, one has the \((d-1)\)-form symmetry of ghost number \(p\) with current \(\Lambda_p\), generalising the ghostly symmetry of Maxwell symmetry with current \(c\) for arbitrary \(p\). Its descendants can be read off from \eqref{eq:abelian_higher_gauge_bv_differential}, and the descent chain stops at \(\Lambda_0\), which is not a symmetry since \(\mathrm d\Lambda_0\) is not \(Q\)-exact.

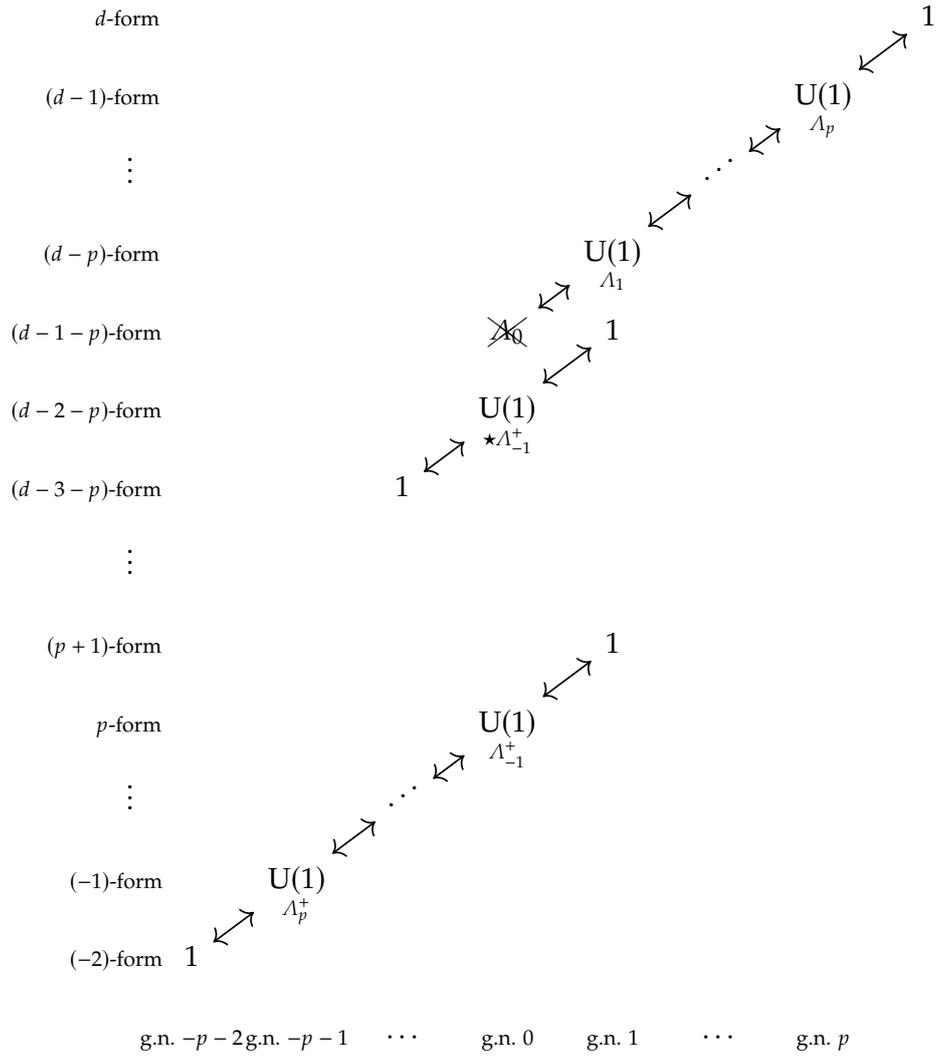
\begin{figure}
\begin{equation*}
    \begin{tikzcd}[row sep={\pgfkeysvalueof{/tikz/commutative diagrams/row sep/large},between origins}, column sep={3.6em,between origins}]
    \scriptstyle\text{\(\hphantom{(\;-0-p)}d\)-form}&&&&&&&&1\dlar[leftrightarrow]\\
    \scriptstyle\text{\(\hphantom{\;-p}(d-1)\)-form}&&&&&&&\underset{\Lambda_p}{\operatorname U(1)}\dlar[leftrightarrow]\\
    \qquad\quad\vdots&&&&&&\iddots\dlar[leftrightarrow]\\
    \scriptstyle\text{\(\hphantom{\;-1}(d-p)\)-form}&&&&&\underset{\Lambda_1}{\operatorname U(1)}\dlar[leftrightarrow]\\
    \scriptstyle\text{\((d-1-p)\)-form}&&&&\xcancel{\Lambda_0}&1\dlar[leftrightarrow]\\
    \scriptstyle\text{\((d-2-p)\)-form}&&&&\underset{\star\Lambda^+_{-1}}{\operatorname U(1)}\dlar[leftrightarrow]\\
    \scriptstyle\text{\((d-3-p)\)-form}&&&1\\
    \qquad\quad\vdots\\
    \scriptstyle\text{\(\hphantom{d-{}}(p+1)\)-form}&&&&&1\dlar[leftrightarrow]\\
    \scriptstyle\text{\(\hphantom{(d-0-{})}p\)-form}&&&&\underset{\Lambda^+_{-1}}{\operatorname U(1)}\dlar[leftrightarrow]\\
    \qquad\quad\vdots&&&\iddots\dlar[leftrightarrow]\\
    \scriptstyle\text{\(\hphantom{\;d-p\;}(-1)\)-form}&&\underset{\Lambda^+_p}{\operatorname U(1)}\dlar[leftrightarrow]\\
    \scriptstyle\text{\(\hphantom{\;d-p\;}(-2)\)-form}&1\\
    &\scriptstyle\text{g.n.\ \(-p-2\)}&\scriptstyle\text{g.n.\ \(-p-1\)}&\cdots&\scriptstyle\text{g.n.\ \(0\)}&\scriptstyle\text{g.n.\ \(1\)}&\cdots&\scriptstyle\text{g.n.\ \(p\)}
    \end{tikzcd}
\end{equation*}
\caption{Ghostly higher symmetries of \(p\)-form Abelian gauge theory}
\label{fig:symmetries_p_form}
\end{figure}

Thus, the higher-form symmetries of \(\operatorname U(1)\) \(p\)-form gauge theory and their generators are given in \cref{fig:symmetries_p_form}.

As for Maxwell theory, the ghostly conservation laws correspond to the numbers of (higher) antiparticle excitations in a Batalin--Vilkovisky-extended Fock space containing excitations of all ghost numbers.

The total complex analysis applies rather succinctly, with the electric total degree $d-p-1$ total form
\begin{equation}
	\alpha^\mathrm{e} 
	= 
	(-)^{\frac12\Upsilon(\Upsilon+1)}\sum_{i=-1}^p(-)^{ip}\Lambda_i^+
\end{equation}
and  the total degree $p$ total form
\begin{equation}
	\alpha
	=
	(-)^{\frac12\Upsilon(\Upsilon+1)}\sum_{i=-1}^{p} \Lambda_i\,.
\end{equation}
For $\alpha^{\mathrm{e}}$, $\mathrm D$-closure holds identically,
$
\mathrm D \alpha^{\mathrm{e}} = 0,
$
and immediately implies the (ghostly) currents and  descent equations given in \eqref{eq:pform_elec_descent}.

For $\alpha$, $\mathrm D$-closure fails,
\begin{equation}
	\mathrm{D}\alpha
	=
	(-)^{\frac12\Upsilon(\Upsilon-1)}\big(\mathrm{d}\Lambda_{-1} - \star^{-1}\Lambda_{-1}^+\big)\,,
\end{equation}
so that imposing $\mathrm D \alpha =0$ reduces the theory to the topological sector of flat connections $B=QX$. Otherwise  $\mathrm D \alpha \neq 0$ yields the magnetic $(d-p-2)$-form symmetry 
\begin{align}
	\mathrm{d}\mathrm{D}\alpha &= Q\big(\mathrm{d}(-)^\Upsilon\alpha\big)
	&
	&\Rightarrow
	&
	0 &= QJ_\mathrm{m}\,,\\
	&&&&\mathrm{d}J_\mathrm{m} &= Q(-\mathrm{d}\Lambda_{-1})\,,\\
	&&&&
	\mathrm{d}(-\mathrm{d}\Lambda_{-1}) &= 0\,.
\end{align}

\subsubsection{Wilson operators}
Similar to the case for Maxwell theory, we can read off the Wilson operators that transform under the ghostly symmetries from the Batalin--Vilkovisky action \eqref{eq:p-form-BV-action}. Thus, due to the term \(\Lambda^+_{i-1}\wedge\mathrm d\Lambda_i\) in \eqref{eq:p-form-BV-action},
for an \((p-i)\)-dimensional boundaryless submanifold \(\Sigma_{p-i}\subset M\),
the Wilson surface
\begin{equation}\exp\mleft(\mathrm i\alpha_{-i}\int_{\Sigma_{p-i}}\Lambda_i\mright)\end{equation}
(with parameter \(\alpha_{-i}\) having ghost number \(-i\))
transforms under the ghostly electric higher-form symmetry with Noether current
\(J_{\mathrm e}^{(i)}=\Lambda^+_{i-1}\)
(except for \(i=p+1\), which is a \((-1)\)-form symmetry and hence has no associated Wilson operators).
Similarly, if \(B=\Lambda^+_{-1}\) (which is \(\mathrm d\)-closed on shell) is locally of the form \(\Lambda^+_{-1}=\mathrm d\tilde A\), then due to the term \(\Lambda^+_{-1}\wedge\mathrm d\tilde A\) in \eqref{eq:p-form-BV-action}, the 't~Hooft operator
\(\int_{\Sigma_{d-p-2}}\tilde A\)
transforms under the magnetic higher-form symmetry with Noether current \(J_{\mathrm m}=\star\Lambda^+_{-1}\).
Finally, due to the term \(\Lambda^+_{i-1}\wedge\mathrm d\Lambda_i\) in \eqref{eq:p-form-BV-action}, the Wilson operator \(\int_{\Sigma_{d-p-1+i}}\Lambda^+_{i-1}\) transforms under the symmetry with Noether current \(\Lambda_i\) for \(1\le i\le p\).

\subsection{Schwarz-type topological gauge theories}

\subsubsection{Abelian Chern--Simons theory}
\begin{figure}
\begin{equation*}
    \begin{tikzcd}[row sep={\pgfkeysvalueof{/tikz/commutative diagrams/row sep/large},between origins}, column sep={4.5em,between origins}]
    \text{\(\hphantom{()-}3\)-form}&&&&&&1\dlar[leftrightarrow]\\
    \text{\(\hphantom{()-}2\)-form}&&&&&\underset c{\operatorname U(1)}\dlar[leftrightarrow]\\
    \text{\(\hphantom{()-}1\)-form}&&&&\underset A{\operatorname U(1)}\dlar[leftrightarrow]\\
    \text{\(\hphantom{()-}0\)-form}&&&\underset{A^+}{\operatorname U(1)}\dlar[leftrightarrow]\\
    \text{\((-1)\)-form}&&\underset{c^+}{\operatorname U(1)}\dlar[leftrightarrow]\\
    \text{\((-2)\)-form}&1\\
    &\text{g.n.\ \(-3\)}&\text{g.n.\ \(-2\)}&\text{g.n.\ \(-1\)}&\text{g.n.\ \(0\)}&\text{g.n.\ \(1\)}&\text{g.n.\ \(2\)}
    \end{tikzcd}
\end{equation*}
\caption{Ghostly symmetries of three-dimensional Abelian Chern--Simons theory with the gauge field \(A\) defined globally. Since this theory is topological, the descent chain never stops.}\label{fig:symmetries_abelian_cs}
\end{figure}
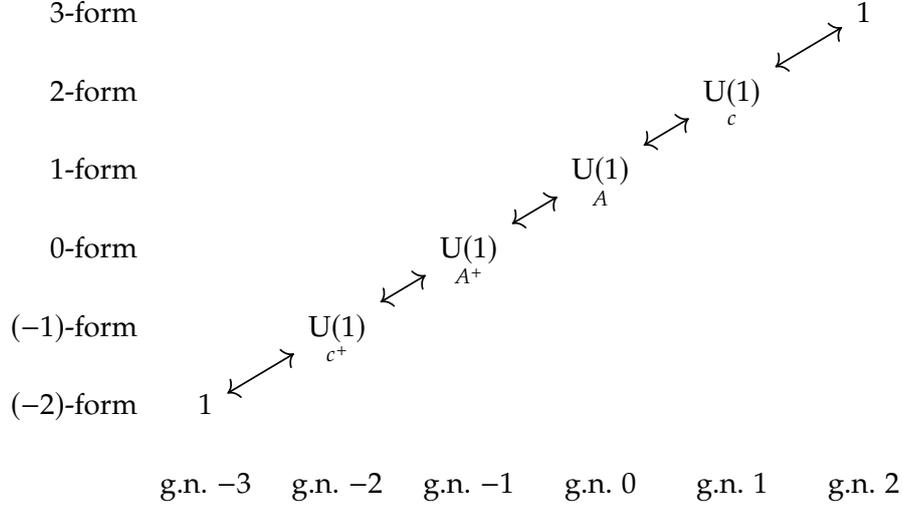
In Maxwell theory, we saw that some descent chains did not continue, as one expects generically for a non-topological theory. By contrast, let us consider Abelian Chern--Simons theory in three dimensions, which is topological; for simplicity we assume that \(A\) is globally defined. The Batalin--Vilkovisky action on a three-dimensional spacetime \(M\) is then \cite{Jurco:2018sby}
\begin{equation}
	S
	=
	\int \tfrac12 A\wedge\mathrm{d}A - \mathrm{d}c\wedge A^+
\end{equation}
where \(c,A,A^+,c^+\) are zero-, one-, two-, three-forms, respectively, with ghost numbers \(1\), \(0\), \(-1\), and \(-2\), respectively. The Batalin--Vilkovisky differential is then
\begin{align}\label{eq:CS_QBV}
	0 &= Qc\,,
	&
	\mathrm{d}c &= Q(-A)\,,
	&
	\mathrm{d}(-A) &= QA^+\,,
	&
	\mathrm{d}A^+ &= Qc^+\,,
	&
	\mathrm{d}c^+ &= 0\,.
\end{align}
Thus, starting from the ghostly two-form symmetry whose Noether current is \(c\), the descent chain does not stop, as expected of a topological field theory. The resulting symmetries are shown in \cref{fig:symmetries_abelian_cs}.

In terms of the cohomology of the total complex,  only $\on{Tot}^1(\mathcal A)$ is non-empty. Thus, we have only one total degree 1 form to consider 
\begin{equation}
	\alpha^\mathrm{CS}
	=
	- c + A + A^+ + c^+
\end{equation}
and $\mathrm D \alpha^{\mathrm{CS}} =0$ imposes precisely \eqref{eq:CS_QBV}. There are no $\mathrm D$-exact  total degree $1$ forms (as there are no total degree $0$ forms) so the only non-empty cohomology is 
\be
\on{H}_{\mathrm D}^1(\on{Tot}^\bullet(\mathcal A))\cong \mathcal \bigoplus_{p,q}\mathcal A^{p,q}
\ee

The (ghostly) higher-form symmetries are characterised by the total complex cohomology, as expected for a topological theory. Note, these all act nontrivially on corresponding Wilson loops (except for the \((-1)\)-form symmetry): \(A\) acts on the Wilson loop \(\int_{\Sigma_1} A\), and \(c\) acts on the Wilson surface \(\int_{\Sigma_2}A^+\), and \(A^+\) acts on the Wilson point \(\int_{\Sigma_0}c\).

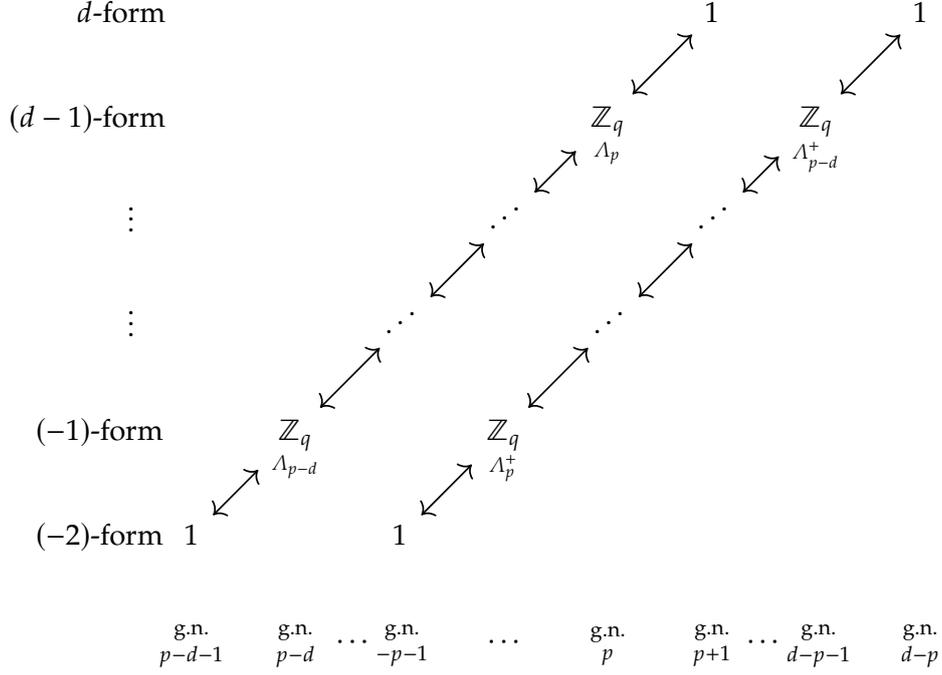
\begin{figure}
\begin{equation*}
    \begin{tikzcd}[row sep={\pgfkeysvalueof{/tikz/commutative diagrams/row sep/huge},between origins}, column sep={3.56em,between origins}]
    \text{\(\hphantom{(0-{})}d\)-form}&&&&&&1\dlar[leftrightarrow]&&1\dlar[leftrightarrow]\\
    \text{\((d-1)\)-form}&&&&&\underset{\Lambda_p}{\mathbb Z_q}\dlar[leftrightarrow]&&\underset{\Lambda^+_{p-d}}{\mathbb Z_q}\dlar[leftrightarrow]\\
    \qquad\quad\vdots&&&&\iddots\dlar[leftrightarrow]&&\iddots\dlar[leftrightarrow]\\
    \qquad\quad\vdots&&&\iddots\dlar[leftrightarrow]&&\iddots\dlar[leftrightarrow]\\
    \text{\(\hphantom{d\;}(-1)\)-form}&&\underset{\Lambda_{p-d}}{\mathbb Z_q}\dlar[leftrightarrow]&&\underset{\Lambda^+_p}{\mathbb Z_q}\dlar[leftrightarrow]\\
    \text{\(\hphantom{d\;}(-2)\)-form}&1&&1\\
    &\genfrac{}{}{0pt}1{\text{g.n.}}{p-d-1}&\genfrac{}{}{0pt}1{\text{g.n.}}{p-d}&\cdots\genfrac{}{}{0pt}1{\text{g.n.}}{-p-1}\hphantom\cdots&\cdots&\genfrac{}{}{0pt}1{\text{g.n.}}{p}&\genfrac{}{}{0pt}1{\text{g.n.}}{p+1}&
    \cdots\genfrac{}{}{0pt}1{\text{g.n.}}{d-p-1}\hphantom\cdots&
    \genfrac{}{}{0pt}1{\text{g.n.}}{d-p}
    \end{tikzcd}
\end{equation*}
\caption{Ghostly symmetries of \(BF\) theory with level \(q\)}\label{fig:symmetries_bf}
\end{figure}

\subsubsection{$BF$ theory}

As another example of an Abelian topological theory, we consider a \(BF\) theory. Suppose that spacetime \(M\) is covered by a sufficiently fine lattice or triangulation so that it makes sense to speak of discrete-valued differential forms (\cref{ssec:discrete_differential_form}). Working with the coefficient ring \(\mathbb Z_q\) of integers modulo \(q\),
we have the action
\begin{equation}
    S=\frac{q}{2\pi}\int_M B\wedge\mathrm dA,
\end{equation}
where \(A\in\Omega^p(M;\mathbb Z_q)\) and \(B\in\Omega^{d-p-1}(M;\mathbb Z_q)\) are potentials.
Both \(A\) and \(B\) can be shifted by exact differential forms; this gauge symmetry produces a tower of ghost fields
\begin{align}
    \Lambda_i&\in\Omega^{p-i}(M;\mathbb Z_q),&\Lambda^+_i&\in\Omega^{d-p+i}(M;\mathbb Z_q), \;(p-d\le i\le p),    
\end{align}
where \(\Lambda_i\) carries ghost number \(i\) while \(\Lambda^+_i\) carries ghost number \(-1-i\), and where we identify \(A=\Lambda_0\) and \(A^+=\Lambda^+_0\) and \(B=\Lambda^+_{-1}\) and \(B^+=\Lambda_{-1}\).
Then \(\Lambda_1,\dotsc,\Lambda_p\) are ghosts for \(A=\Lambda_0\) while \(\Lambda^+_{-2},\dotsc,\Lambda^+_{p-d}\) are ghosts for \(B=\Lambda^+_{-1}\) and the corresponding antifield ghosts are  \(\Lambda^+_{-1},\dotsc,\Lambda^+_{p}\) and  \(\Lambda_{-2},\dotsc,\Lambda_{p-d}\), respectively. 

The Batalin--Vilkovisky action is then
\begin{equation}
    S=\frac{q}{2\pi}\int_M
    \Lambda^+_{p-d}\wedge\mathrm d\Lambda_{p-d+1}
    +\dotsb
    +\Lambda^+_{p-1}\wedge\mathrm d\Lambda_p.
\end{equation}
The Batalin--Vilkovisky algebra is then
\begin{equation}
    Q\Lambda_i=\begin{cases}
        \mathrm d\Lambda_{i+1}&(i\le p-1)\\
        0&(i=p),
    \end{cases}
\end{equation}
which are already the descent equations.
If we assume that all potentials are globally defined, then 
ghostly higher symmetries are as given in \cref{fig:symmetries_bf}. In particular, at ghost number \(0\), we have a \(p\)-form symmetry (whose current is \(B\)) and a \((d-1-p)\)-form symmetry (whose current is \(A\)), as noted in e.g.\ \cite{Nguyen:2024ikq} for \(d=3\), \(p=1\).

\subsection{Coupling to matter and ultraviolet--infrared matching}\label{ssec:maxwell_matter}
It is well known \cite{Gaiotto:2014kfa} that for the electric \(\operatorname U(1)\)-valued one-form symmetry, when the Maxwell field is coupled to matter of charge \(q\), then the \(\operatorname U(1)\) breaks into a \(\mathbb Z_q\) subgroup, and that this symmetry can be matched with that of the effective theory in the deep infrared, which is a \(\mathbb Z_q\)-valued \(BF\) theory. In this way, one can match the symmetries of the infrared and ultraviolet theories similar to 't~Hooft anomaly matching.
In this section, we demonstrate in the case of matter coupling to Maxwell theory that the symmetry matching also holds for ghostly symmetries.

We start by coupling a complex scalar field \(\Phi\) of charge \(q\in\mathbb Z\) to Maxwell theory \eqref{eq:maxwell-action} through the standard minimal-coupling 
\begin{align}\label{eq:matter-coupling}
	&\int \mathrm{d}_A\Phi\wedge\star\mathrm{d}_A\Phi^\ast + 2\pi iq c\big(\Phi\Phi^+ - \Phi^\ast\Phi^{\ast+}\big)\,,
	&
	&
	\begin{aligned}
		\mathrm{d}_A\Phi\phantom{^\ast} &= (\mathrm{d} + 2\pi \mathrm{i}qA)\Phi
		\\
		\mathrm{d}_A\Phi^\ast &= (\mathrm{d} - 2\pi \mathrm{i}qA)\Phi^\ast
	\end{aligned}\,,
\end{align}
where $\Phi^+$ is the scalar antifield and  under the $\on U(1)$ gauge symmetry
\be
\Phi\mapsto \on e^{\on i2\pi q \alpha}\Phi\,,\quad  \Phi^+\mapsto \on  e^{-\on i2\pi q \alpha}\Phi^+\,.
\ee
Then \(QA^+\)  in the \eqref{eq:Max_QBV} picks-up an additional term 
\begin{equation}\label{eq:Max_QBV_complex_scalar}
	QA^+ = -\mathrm{d}B - J_\Phi\,,
\end{equation}
where 
\begin{equation}
	J_\Phi 
	= \frac{\delta S_\mathrm{matter}}{\delta A} 
	=
	2\pi \textrm{i} q\star\big(\Phi\mathrm{d}_A\Phi^\ast - \Phi^\ast\mathrm{d}_A\Phi\big)
	=
	-4\pi q\star\operatorname{Im}(\Phi\mathrm{d}_A\Phi^\ast) 
\end{equation}
is the usual gauge-invariant conserved current with \(\mathrm d_A=\mathrm d-2\pi q\mathrm iA\).

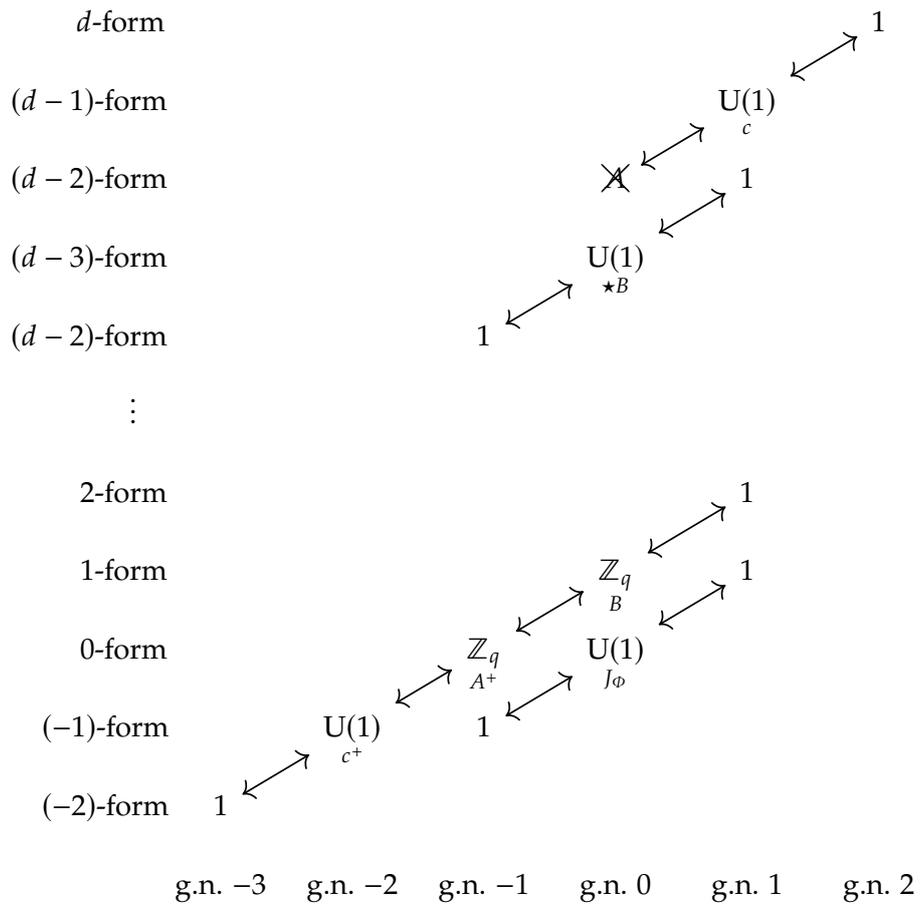
\begin{figure}
\begin{equation*}
    \begin{tikzcd}[row sep={\pgfkeysvalueof{/tikz/commutative diagrams/row sep/large},between origins}, column sep={4.5em,between origins}]
    \text{\(\hphantom{(\;-0)}d\)-form}&&&&&&1\dlar[leftrightarrow]\\
    \text{\((d-1)\)-form}&&&&&\underset c{\operatorname U(1)}\dlar[leftrightarrow]\\
    \text{\((d-2)\)-form}&&&&\xcancel{A}&1\dlar[leftrightarrow]\\
    \text{\((d-3)\)-form}&&&&\underset{\star B}{\operatorname U(1)}\dlar[leftrightarrow]\\
    \text{\((d-2)\)-form}&&&1\\
    \qquad\quad\vdots\\
    \text{\(\hphantom{()-d}2\)-form}&&&&&1\dlar[leftrightarrow]\\
    \text{\(\hphantom{()-d}1\)-form}&&&&\underset B{\mathbb Z_q}\dlar[leftrightarrow]&1\dlar[leftrightarrow]\\
    \text{\(\hphantom{()-d}0\)-form}&&&\underset{A^+}{\mathbb Z_q}\dlar[leftrightarrow]&\underset{J_\Phi}{\operatorname U(1)}\dlar[leftrightarrow]\\
    \text{\(\hphantom{\;d\;}(-1)\)-form}&&\underset{c^+}{\operatorname U(1)}\dlar[leftrightarrow]&1\\
    \text{\(\hphantom{\;d\;}(-2)\)-form}&1\\
    &\text{g.n.\ \(-3\)}&\text{g.n.\ \(-2\)}&\text{g.n.\ \(-1\)}&\text{g.n.\ \(0\)}&\text{g.n.\ \(1\)}&\text{g.n.\ \(2\)}
    \end{tikzcd}
\end{equation*}
\caption{Ghostly higher-form symmetries of Maxwell theory in \(d\) spacetime dimensions coupled to charged matter of charged \(q\)}\label{fig:symmetries_maxwell_matter}
\end{figure}

Consequently, the corresponding would-be descent equation in \eqref{eq:maxwell_electric_descent} is no longer exact  so the \(\operatorname U(1)^{[0,-1]}\) symmetry must break. Since $B$ and $J_\Phi$ are subject to the semi-classical quantisation conditions
\be
\int_{\Sigma_{d-1}} H \in 2\pi \mathrm i \mathbb{Z}, \qquad \int_{\Sigma_{d-1}} J_\Phi \in 2\pi q\mathrm i \mathbb{Z}
\ee
where locally $H=\mathrm d B$  and $\Sigma_{d-1}$ is a homology $(d-1)$-cycle, we observe that the  discrete subgroup  \(\mathbb Z_q^{[0,-1]}\) survives, as shown in \cref{fig:symmetries_maxwell_matter}. 

Of course, now that we have matter, there is the usual conserved current \(J_\Phi\), which comes with its own descendants. As per the general discussion  given in \cref{ssec:int_coho}, writing \(j_\Phi = \star^{-1}J_\Phi\), we see that while the would-be descent equation $QA^+=\mathrm dB$ fails to hold  identically,  from \eqref{eq:Max_QBV_complex_scalar} we inherit 
\begin{equation}
    Q(\mathrm d A^+) = -\mathrm d(\star j_\Phi),
\end{equation}
which yields, in precise analogy to the  $\star B$ current of the magnetic $(d-3)$-form symmetry   of pure Maxwell theory, the descent equations 
\begin{align}
	0 &= Q(\star j_\Phi)\,,
	&
	\mathrm{d}(\star j_\Phi) &= Q(-\mathrm{d}A^+)\,,
	&
	\mathrm{d}(-\mathrm{d}A^+) &= 0\,,
\end{align}
where the current $-\mathrm d A^+$ is trivial ($\mathrm d$-exact).

\subsubsection{Matching with the infrared theory}
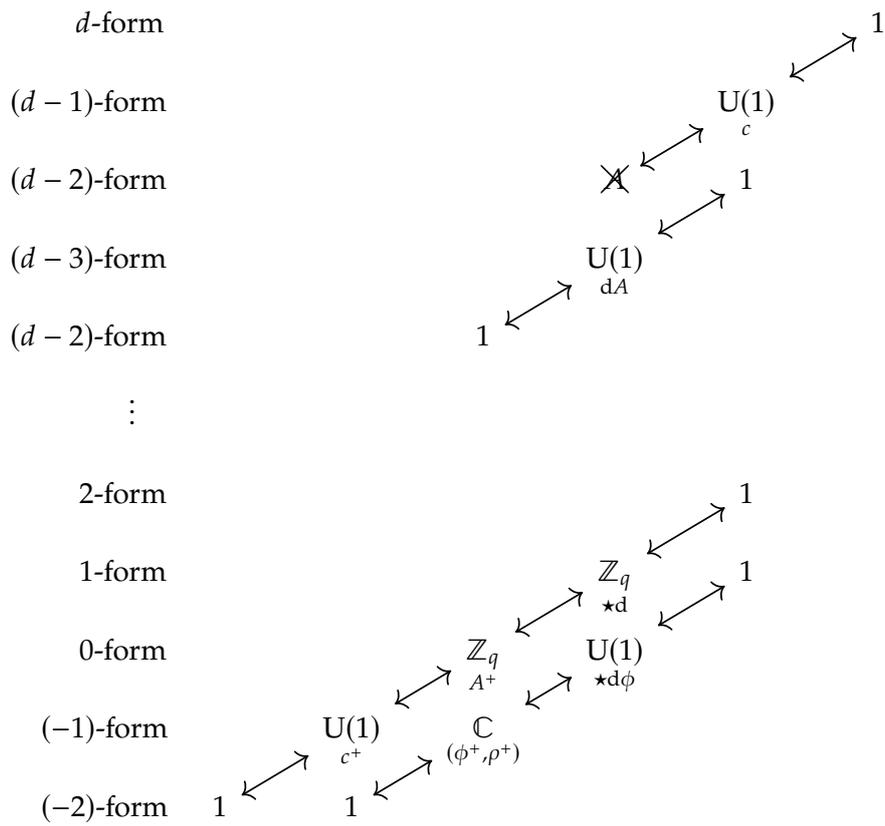
\begin{figure}
\begin{equation*}
    \begin{tikzcd}[row sep={\pgfkeysvalueof{/tikz/commutative diagrams/row sep/large},between origins}, column sep={4.5em,between origins}]
    \text{\(\hphantom{(\;-0)}d\)-form}&&&&&&1\dlar[leftrightarrow]\\
    \text{\((d-1)\)-form}&&&&&\underset c{\operatorname U(1)}\dlar[leftrightarrow]\\
    \text{\((d-2)\)-form}&&&&\xcancel{A}&1\dlar[leftrightarrow]\\
    \text{\((d-3)\)-form}&&&&\underset{\mathrm dA}{\operatorname U(1)}\dlar[leftrightarrow]\\
    \text{\((d-2)\)-form}&&&1\\
    \qquad\quad\vdots\\
    \text{\(\hphantom{()-d}2\)-form}&&&&&1\dlar[leftrightarrow]\\
    \text{\(\hphantom{()-d}1\)-form}&&&&\underset{\star\mathrm d}{\mathbb Z_q}\dlar[leftrightarrow]&1\dlar[leftrightarrow]\\
    \text{\(\hphantom{()-d}0\)-form}&&&\underset{A^+}{\mathbb Z_q}\dlar[leftrightarrow]&\underset{\star\mathrm d\phi}{\operatorname U(1)}\dlar[leftrightarrow]\\
    \text{\(\hphantom{\;d\;}(-1)\)-form}&&\underset{c^+}{\operatorname U(1)}\dlar[leftrightarrow]&\underset{(\phi^+,\rho^+)}{\mathbb C}\dlar[leftrightarrow]\\
    \text{\(\hphantom{\;d\;}(-2)\)-form}&1&1\\
    \end{tikzcd}
\end{equation*}
\caption{Ghostly higher-form symmetries of the infrared limit of Maxwell theory in \(d\) spacetime dimensions coupled to charged matter of charged \(q\), which agree with those of the ultraviolet theory given in \cref{fig:symmetries_maxwell_matter}}\label{fig:symmetries_maxwell_matter_infrared}
\end{figure}
We now check whether these symmetries persist in the infrared.
To obtain the infrared theory, we decompose \(\Phi\) into radial and angular parts:
\begin{equation}
	\Phi = \tfrac{1}{\sqrt2}e^{2\pi\mathrm{i} \phi}\rho
\end{equation}

Then \eqref{eq:matter-coupling} becomes
\begin{equation}\label{eq:stueckelberg}
	\int \tfrac12 \mathrm{d}\rho\wedge\star\mathrm{d}\rho + \tfrac12 (2\pi\rho)^2 (\mathrm{d}\phi + qA)\wedge\star(\mathrm{d}\phi + qA) + qc\phi^+
\end{equation}
where we have identified
\begin{equation}
	\phi^+ 
	=
	2\pi \mathrm{i}\big(\Phi\Phi^+ - \Phi^\ast\Phi^{\ast +}\big)
\end{equation}
to obtain the correct gauge transformation
\begin{align}
	A &\to A - \mathrm{d}\alpha
	&
	\phi &\to \phi + q\alpha
\end{align}
This now appears as a Stueckelberg coupling (see \cite{Ruegg:2003ps} for a review) since \(\phi\colon M\to\operatorname U(1)=\mathbb R/\mathbb Z\) transforms affinely as a principal homogeneous space (i.e.\ \(\operatorname U(1)\)-torsor).
Now, in the deep infrared,
we may set \(\rho\) to a vacuum expectation value \(v\) and discard  the suppressed higher-derivative terms  to obtain the model
\begin{equation}
	S_{\mathrm{IR}}[A,c,A^+,c^+,\phi,\phi^+]
	=
	\int (2\pi v)^2qA\wedge\star\mathrm{d}\phi + qc\phi^+ - \mathrm{d}c\wedge A^+.
\end{equation}
(Note that we do \emph{not} dualise \(\phi\) to obtain a \(BF\) model since, as we saw above, dualisation changes the ghostly symmetries.)
Now \(A\) and \(\phi\) can be taken to be a \(\mathbb Z_q\)-valued one-form and a \(\mathbb Z_q\)-valued zero-form respectively.
The non-ghostly electric symmetry \(J_\mathrm e=-{\star^{-1}}\mathrm dA\) and magnetic symmetry \(J_\mathrm m=\mathrm dA\) are well known to persist in the infrared as \(\mathbb Z_q\)-valued symmetries. (The field \(A\) does \emph{not} form a current, even though it is closed on shell, since \(A\) is only locally defined.)
Similarly, \(A^+\) is now a \(\mathbb Z_q\)-valued \((d-1)\)-form, forming the current for the ghostly symmetry \(\mathbb Z_q^{[0,-1]}\).

The ghostly symmetries \(\operatorname U(1)^{[0,-1]}\) and \(\operatorname U(1)^{[-1,-2]}\) also persist since
\begin{align}
    J_{\mathrm e}^{(1)}&=A^+,&
    J_{\mathrm e}^{(2)}&=c^+
\end{align}
are still \(\mathrm d\)-closed up to \(Q\)-exact terms:
\begin{align}
    \mathrm dA^+&=Qc^+,&
    \mathrm dc^+&=0.
\end{align}
Finally, \(v^2\star \mathrm d\phi\) is the corresponding current, is still conserved; its descendant is \(\phi^+\), which combined with \(\rho^+\) fills out the \(\mathbb C\)-valued \((-1)\)-form ghostly symmetry.
Thus the infrared set of symmetries given in \cref{fig:symmetries_maxwell_matter_infrared}, both ghostly and non-ghostly, match those of \cref{fig:symmetries_maxwell_matter} in the ultraviolet.

\subsubsection{Coupling matter to ghostly gauge symmetry}
\begin{figure}
\begin{equation*}
    \begin{tikzcd}[row sep={\pgfkeysvalueof{/tikz/commutative diagrams/row sep/huge},between origins}, column sep={\pgfkeysvalueof{/tikz/commutative diagrams/column sep/huge},between origins}]
    \qquad\quad\vdots&&&&&&\iddots\dlar[leftrightarrow]\\    
    \text{\(\hphantom+2\)-form}&&&&&\underset{\Lambda^+_{p-3}}{\operatorname U(1)}\dlar[leftrightarrow]\\
    \text{\(\hphantom+1\)-form}&&&&\underset{\Lambda^+_{p-2}}{\mathbb Z_q}\dlar[leftrightarrow]&1\dlar[leftrightarrow]\\
    \text{\(\hphantom+0\)-form}&&&\underset{\Lambda^+_{p-1}}{\mathbb Z_q}\dlar[leftrightarrow]&\underset{J_\rho}{\operatorname U(1)}\dlar[leftrightarrow]\\
    \text{\((-1)\)-form}&&\underset{\Lambda^+_p}{\operatorname U(1)}\dlar[leftrightarrow]&\underset{\rho^+}{\operatorname U(1)}\dlar[leftrightarrow]\\
    \text{\((-2)\)-form}&1&1\\
    &\genfrac{}{}{0pt}0{\text{g.n.}}{-2-p}&\genfrac{}{}{0pt}0{\text{g.n.}}{-1-p}&\genfrac{}{}{0pt}0{\text{g.n.}}{-p}&\genfrac{}{}{0pt}0{\text{g.n.}}{1-p}&\genfrac{}{}{0pt}0{\text{g.n.}}{2-p}&\cdots
    \end{tikzcd}
\end{equation*}
\caption{Ghostly higher symmetries of \(p\)-form Abelian gauge theory with sources of charge \(q\) coupled to the one-form ghost \(\Lambda_{p-1}\) (only lower form degrees shown)}
\label{fig:symmetries_p_form_matter}
\end{figure}
Abelian higher gauge theories (\cref{ssec:higher_abelian_gauge_theory}) have a one-form field \(\Lambda_{p-1}\), but unlike the photon in Maxwell theory this field has ghost number \(p-1\) (Maxwell theory corresponds to the case \(p=1\)).
We can generalise the previous discussion (\cref{ssec:maxwell_matter}) of coupling matter of charge \(q\) to the Maxwell theory's one-form by
coupling `matter' of charge \(q\) and ghost number \(1-p\) to the one-form field \(\Lambda_{p-1}\) of Abelian higher gauge theory.

It may initially seem strange to couple `matter' of nonzero ghost number to ghost fields.
However, such phenomena can occur in twisted theories (where, due to redefinition of ghost number during twisting, matter that used to have zero ghost number may have nonzero ghost number afterwards); more generally, in the Batalin--Vilkovisky formalism, during the construction of the one-particle-irreducible effective action, it is natural to couple sources to fields regardless of their ghost number, including for ghosts and antifields (see \cite{Batalin:1981jr} for sources for ghosts, and  \cite{Batalin:2013xpa,Borsten:2025hrn} for sources for the antifields).

An apparent obstacle to introducing charged `matter' to ghostly symmetries is the nonzero ghost number: writing down a would-be covariant derivative
\begin{equation}
    (\mathrm d-2\pi\mathrm iq\Lambda_{p-1})\Phi
\end{equation}
for a complex scalar field \(\Phi\) is problematic since \((\mathrm d-2\pi\mathrm iq\Lambda_{p-1})\Phi\) would not have homogeneous ghost number.
Instead, one must directly work with the Stueckelberg form \eqref{eq:stueckelberg}, where the `angular component' \(\phi\) now has homogeneous ghost number \(1-p\).
Then the rest of the discussion of \cref{ssec:maxwell_matter} goes through mutatis mutandis,
resulting in the symmetries shown in \cref{fig:symmetries_p_form_matter}.

\section{Operators in cohomological topological field theories}\label{sec:tqft}
A cohomological (or Witten-type) topological field theory (TQFT)
is a topologically protected subsector of a supersymmetric field theory, obtained by a procedure known as \emph{twisting}.
From the perspective of the Batalin--Vilkovisky formalism,
twisting may be thought of as a redefinition of the Batalin--Vilkovisky differential as
\begin{equation}
    Q_\mathrm{new}\coloneqq Q_\mathrm{old}+Q',
\end{equation}
where the Batalin--Vilkovisky differential of the twisted theory \(Q_\mathrm{new}\) is
a sum of the original Batalin--Vilkovisky differential \(Q_\mathrm{old}\) and
\(Q'\), which is one of the spacetime supersymmetry generators.
This redefinition is accompanied by a redefinition of the ghost number such that the new differential \(Q_\mathrm{new}\) is homogeneously of ghost number \(1\) (whereas prior to twisting only \(Q_\mathrm{old}\) had ghost number \(1\), with the supersymmetry \(Q'\) being of ghost number \(0\)), as well as a redefinition of the Lorentz group so as to make it commute with \(Q_\mathrm{new}\).
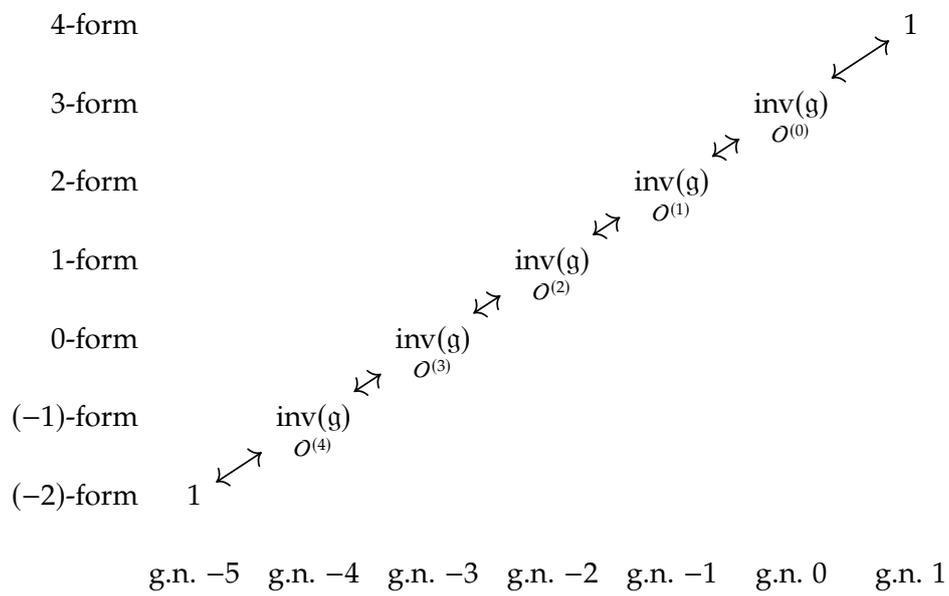
\begin{figure}
\begin{equation*}
    \begin{tikzcd}[row sep={\pgfkeysvalueof{/tikz/commutative diagrams/row sep/large},between origins}, column sep={4.083em,between origins}]
    \text{\(\hphantom{()-}4\)-form}&&&&&&&1\dlar[leftrightarrow]\\
    \text{\(\hphantom{()-}3\)-form}&&&&&&\underset{\mathcal O^{(0)}}{\operatorname{inv}(\mathfrak g)}\dlar[leftrightarrow]\\
    \text{\(\hphantom{()-}2\)-form}&&&&&\underset{\mathcal O^{(1)}}{\operatorname{inv}(\mathfrak g)}\dlar[leftrightarrow]\\
    \text{\(\hphantom{()-}1\)-form}&&&&\underset{\mathcal O^{(2)}}{\operatorname{inv}(\mathfrak g)}\dlar[leftrightarrow]\\
    \text{\(\hphantom{()-}0\)-form}&&&\underset{\mathcal O^{(3)}}{\operatorname{inv}(\mathfrak g)}\dlar[leftrightarrow]\\
    \text{\((-1)\)-form}&&\underset{\mathcal O^{(4)}}{\operatorname{inv}(\mathfrak g)}\dlar[leftrightarrow]\\
    \text{\((-2)\)-form}&1\\
    &\text{g.n.\ \(-5\)}&\text{g.n.\ \(-4\)}&\text{g.n.\ \(-3\)}&\text{g.n.\ \(-2\)}&\text{g.n.\ \(-1\)}&\text{g.n.\ \(0\)}&\text{g.n.\ \(1\)}
    \end{tikzcd}
\end{equation*}
\caption{Ghostly symmetries corresponding to a descent chain of topological operators in Donaldson--Witten theory with gauge Lie algebra \(\mathfrak g\).
The vector space \(\operatorname{inv}(\mathfrak g)\) of invariant polynomials of \(\mathfrak g\) is regarded as an additive Abelian group.
}\label{fig:symmetries_donaldson_witten}
\end{figure}
In the context of TQFTs, the descent equation has been known for a long time since the work of Witten \cite{Witten:1988ze}, where starting from a gauge-invariant topological operator one produces a chain of differential-form topological operators of higher form degree.
These operators fit the definition of a higher-form symmetry perfectly in that they are topological (i.e.\ \(\mathrm d\)-closed up to \(Q\)-exact terms but not \(\mathrm d\)-exact) and physical (i.e.\ gauge-invariant, viz.\ \(Q\)-closed up to \(\mathrm d\)-exact terms, and not vanishing on shell, viz.\ not itself \(Q\)-exact).
As a well known example,
consider the Donaldson--Witten theory \cite{Witten:1988ze} (reviewed in \cite{Labastida:1997pb,Labastida:2005zz}), which is the topological twist of a four-dimensional \(\mathcal N=2\) supersymmetric gauge theory with only vector multiplets, with a gauge group \(G\) (whose Lie algebra is \(\mathfrak g\)).
On a spacetime four-manifold \(M\), the twisted field content is then as follows:
\begin{equation}
\begin{aligned}
    c&\in\Omega^0(M;\mathfrak g),&\phi&\in\Omega^0(M;\mathfrak g),&
    \bar\phi&\in\Omega^0(M;\mathfrak g),&
    A&\in\Omega^1(M;\mathfrak g),\\
    \chi&\in\Omega^2_+(M;\mathfrak g),&
    \eta&\in\Omega^0(M;\mathfrak g),&
    D&\in\Omega^2_+(M;\mathfrak g),&
    \psi&\in\Omega^1(M;\mathfrak g),\\
    c^+&\in\Omega^4(M;\mathfrak g),&\phi^+&\in\Omega^4(M;\mathfrak g),&
    \bar\phi^+&\in\Omega^d(M;\mathfrak g),&
    A^+&\in\Omega^3(M;\mathfrak g),\\
    \chi^+&\in\Omega^2_+(M;\mathfrak g),&
    \eta^+&\in\Omega^d(M;\mathfrak g),&
    D^+&\in\Omega^2_+(M;\mathfrak g),&
    \psi^+&\in\Omega^1(M;\mathfrak g),
\end{aligned}
\end{equation}
where \(\Omega^2_+\) denotes the space of self-dual two-forms.
The fields have ghost numbers \(1\), \(-2\), \(2\), \(0\), \(-1\), \(-1\), \(0\), \(1\) respectively, and the antifields have complementary ghost numbers
\(-2\), \(1\), \(-3\), \(-1\), \(0\), \(0\), \(-1\), \(-2\) respectively. Note, all fields are now (ghost number $q$) differential-form operators in $\mathcal{A}^{p,q}$; the twisted theory can be formulated on an arbitrary manifold, while its  untwisted  parent requires vanishing first and second Stiefel--Whitney class.
We omit the detailed Batalin--Vilkovisky transformations, which are the action of \(Q'\) (as found in e.g.\ \cite{Labastida:1997pb}) plus the standard gauge-transformation terms.

Consider the space \(\operatorname{inv}(\mathfrak g)\) of invariant polynomials of the Lie algebra \(\mathfrak g\). For any invariant polynomial \(p\in\operatorname{inv}(\mathfrak g)\), then, the operator \(\mathcal O^{(0)}\coloneqq p(\phi)\in\mathcal A^{0,0}\) is a gauge-invariant zero-form operator with ghost number zero, so one can use it to start a descent chain:
\begin{equation}
\begin{aligned}
    0&=Q\mathcal O^{(0)},&
    \mathrm d\mathcal O^{(0)}&=Q\mathcal O^{(1)},&
    \mathrm d\mathcal O^{(1)}&=Q\mathcal O^{(2)},\\
    \mathrm d\mathcal O^{(2)}&=Q\mathcal O^{(3)},&
    \mathrm d\mathcal O^{(3)}&=Q\mathcal O^{(4)},&
    \mathrm d\mathcal O^{(4)}&=0.
\end{aligned}
\end{equation}
The fact that the theory is topological ensures that the descent chain continues until it vanishes due to form-degree reasons.
For example, for \(\mathfrak g=\mathfrak{su}(2)\), there is only one independent  invariant polynomial (the Killing form), and the operators are \cite{Witten:1988ze}:
\begin{equation}
\begin{aligned}
    \mathcal{O}^{(0)}&=\operatorname{tr}(\phi^2),&
    \mathcal{O}^{(1)}&=\operatorname{tr}(\phi\psi),&
    \mathcal{O}^{(2)}&=\operatorname{tr}\left(\frac12\psi\wedge\psi+\mathrm i\phi\wedge F\right),\\
    \mathcal{O}^{(3)}&=\mathrm i\operatorname{tr}(\psi\wedge F),&
    \mathcal{O}^{(4)}&=-\frac12\operatorname{tr}(F\wedge F).
\end{aligned}
\end{equation}
{Note that the ghost $c$ contributions to $Q\mathcal{O}^{(i)}$ drop-out as they are gauge-invariant}. 

{Then,  for $\gamma_i \in \on H_i(X)$, the operators
\begin{equation}
W_{\mathcal{O}^{(i)}}^{\left(\gamma_i\right)}=\int_{\gamma_i} \mathcal{O}^{(i)}
\end{equation}
are  topological observable and therefore each $\mathcal{O}^{(i)}$ corresponds to a  (ghostly) $(d-i-1)$-form symmetry.} Namely, the $\mathcal{O}^{(i)}$ constitute Noether currents for ghostly three-, two-, one-, zero, and \((-1)\)-form symmetries as shown in \cref{fig:symmetries_donaldson_witten}.

 If these are to be regarded as symmetries, what operators transform under them? The action of Donaldson--Witten theory does not contain any terms of the form
\begin{equation}\label{eq:tqft_coupling}
    \mathcal O^{(i)}\wedge\mathrm dW^{(i)}
\end{equation}
However, one can always couple this topological theory to another (possibly non-topological) theory \(\mathcal T'\) by introducing terms of the form \eqref{eq:tqft_coupling} where now \(W^{(i)}\) are differential \((4-i)\)-forms with ghost number \(i\) built out of fields of \(\mathcal T'\).
Then the Wilson loops \(\int_{\Sigma_{4-i}} W^{(i)}\) transform nontrivially under the ghostly symmetry associated to the Noether current \(\mathcal O^{(i)}\).

{The obvious implication is that the Donaldson invariants, as understood from the point of view of TQFT via Donaldson--Witten theory, are correlation functions of higher--form conserved charges related by descent equations. This raises the question of what, if any, analogous role the correlation functions of the (ghostly) higher-form conserved charges play.}

\section{Centre symmetry of Yang--Mills theory and its descendants}\label{sec:centre}
Pure Yang--Mills theory with a simple gauge group \(G\) has a discrete non-ghostly one-form symmetry valued in the centre \(\operatorname Z(G)\), as discovered by Polyakov and 't~Hooft \cite{Polyakov:1976fu,Polyakov:1975rs,tHooft:1977nqb}.\footnote{Of course, since this work predates the present understanding of generalised global symmetries, Polyakov and 't~Hooft  did not articulate the Yang--Mills centre symmetry in terms of a one-form symmetry.}
This centre one-form symmetry plays a crucial role in confinement, as reviewed in \cite{Ogilvie:2012is,Holland:2000uj}:
at finite temperature (viz.\ with periodic Euclidean time), a timelike Wilson loop (or Polyakov loop, in this context) transforms nontrivially under the centre symmetry and serves as an order parameter for confinement.
This section shows that the centre symmetry of pure Yang--Mills theory begets ghostly zero-form and \((-1)\)-form symmetries by presenting it in a form where a magnetic \(\operatorname Z(G)\) symmetry is explicitly gauged; this may be thought of as a continuum non-Abelian analogue of the Villain formulation \cite{Villain:1974ir} of lattice \(\operatorname U(1)\) gauge theory.

For simplicity, we assume \(G\) be a simply connected compact simple Lie group with centre \(\operatorname Z(G)\), such that the corresponding centreless Lie group is \(\tilde G\coloneqq G/\operatorname Z(G)\) with fundamental group \(\pi_1(\tilde G)=\operatorname Z(G)\).
It is well known that pure Yang--Mills theory with gauge group \(G\) enjoys an electric \(\operatorname Z(G)\)-valued one-form symmetry, while the magnetic \((d-3)\)-form symmetry is trivial since $\pi_1(G)=1$. On the other hand   pure Yang--Mills theory with gauge group \(\tilde G\) enjoys a magnetic \(\pi_1(\tilde G)\)-valued magnetic \((d-3)\)-form symmetry, while the electric \(1\)-form symmetry is trivial since $Z(\tilde G)=1$.

 The former is `electric' insofar as it can be broken by coupling to electrically charged matter, while the latter cannot --- the same way that, for Maxwell theory, the electric one-form symmetry with Noether current \(\star F\) can be broken by coupling to electrically charged matter but the magnetic \((d-3)\)-form symmetry cannot. For Maxwell theory (\cref{ssec:maxwell}), the electric symmetry admits descendants whereas the magnetic symmetry does not, as the first putative descendant is exact. We here show that an analogous situation holds in the case of a Villain-like formulation of Yang--Mills theory.

\subsection{Čech cocycles}
Since the centre symmetry is discrete, we must work with a model of cohomology that works with discrete coefficients, unlike the de~Rham cohomology of differential forms, such as cellular cohomology, simplicial cohomology or Čech cohomology.  We adopt Čech cohomology since, for Yang--Mills theory, it is  the  familiar   patchwise description of the gauge field where the connections on intersections  are related by gauge transformations. (For a review of Čech cohomology, see \cite{Nakahara:2003nw}.)

Suppose that spacetime \(M=\bigcup_iU_i\) is covered by a sufficiently fine open cover \(\{U_i\}_{i\in I}\).
Let \(\Gamma\) be a Lie group that will serve as the coefficients; we use multiplicative notation for \(\Gamma\), even when \(\Gamma\) is Abelian.
A \emph{\(\Gamma\)-valued \(p\)-cochain \(\alpha\)} is a collection of smooth maps
\begin{equation}
    \alpha_{i_0\dotso i_p}\colon U_{i_0}\cap\dotsb\cap U_{i_p}\to \Gamma
\end{equation}
for each ordered \((p+1)\)-tuple of open sets \(U_{i_0},\dotsc,U_{i_p}\) that is totally graded-antisymmetric with respect to permutation of the indices, i.e.
\begin{equation}
    \alpha_{i_{\sigma(0)}\dotso i_{\sigma(p)}}
    =\alpha^{(-1)^\sigma}_{i_0\dotso i_p},
\end{equation}
where \((-1)^\sigma\in\{\pm1\}\) is the parity of the permutation \(\sigma\colon\{0,\dotsc,p\}\to\{0,\dotsc,p\}\).
Let us call the collection of \(\Gamma\)-valued \(p\)-cochains \(\vOmega{}^p(U;\Gamma)\), mimicking the notation for differential forms.

If \(\Gamma\) is Abelian, one can define the differential
\begin{equation}
\begin{split}
    \mathrm d\colon\vOmega{}^p(U;\Gamma)&\to\vOmega{}^{p+1}(U;\Gamma),\\
    (\mathrm d\alpha)_{i_0\dotso i_{p+1}}(x)&\coloneqq\prod_{j=0}^{p+1}\left(\alpha_{i_0\dotso\hat i_j\dotso i_{p+1}}(x)\right)^{(-1)^{p-j}},
\end{split}
\end{equation}
where the circumflex \(\hat i_j\) means that the \(j\)th index \(i_j\) is to be omitted. One then shows that \(\mathrm d^2=0\), and the corresponding cohomology defines Čech cohomology. Furthermore, if \(\Gamma\) in fact forms a ring rather than just an additive Abelian group, then there exists a wedge product of Čech cochains that descends to the cup product of Čech cohomology classes. Thus \(\vOmega{}^\bullet(U;\Gamma)\) serves as a discrete analogue of \(\Gamma\)-valued differential forms, and we will speak of \(\Gamma\)-valued differental forms as synonyms of \(\Gamma\)-valued cochains whenever convenient.

If \(\Gamma\) is non-Abelian, the spaces \(\vOmega{}^p(U;\Gamma)\) are still well defined, but the definition of \(\mathrm d\colon\vOmega{}^p(U;\Gamma)\to\vOmega{}^{p+1}(U;\Gamma)\) is ambiguous when \(p\ge0\) due to ordering issues. For \(\vOmega{}^0(U;\Gamma)\to\vOmega{}^1(U;\Gamma)\) and \(\vOmega{}^1(U;\Gamma)\to\vOmega{}^2(U;\Gamma)\), however, we may pick the following canonical orderings\footnote{The data for $2$-cocycles and beyond requires higher structures \cite{wedhorn}. For instance the $2$-cocycle data constitutes a $2$-group.}:
\begin{align}
    (\mathrm d\alpha)_{ij}&\coloneqq \alpha_i\alpha_j^{-1},&
    (\mathrm d\alpha)_{ijk} &\coloneqq \alpha_{ij}\alpha_{jk}\alpha_{ik}^{-1}.
\end{align}
Then one checks that indeed \(\mathrm d^2\colon\vOmega{}^0(U;\Gamma)\to\vOmega{}^2(U;\Gamma)\) vanishes, and one can define non-Abelian cohomology in degree one as
\begin{equation}
    \operatorname H^1(U;G)
    = \left\{\alpha\in\vOmega{}^2(U;G)\middle|\mathrm d\alpha=1\right\} / \sim
\end{equation}
(we write \(\mathrm d\alpha=1\) rather than \(\mathrm d\alpha=0\) to stick to multiplicative notation for \(\Gamma\)), where the equivalence relation \(\sim\) is defined as
\begin{equation}
    g\sim g'\iff \exists h\in\vOmega{}^1(U;G)\colon g'_{ij}=h_ig_{ij}h_j^{-1}.
\end{equation}
This non-Abelian cohomology then classifies principal \(\Gamma\)-bundles \cite{wedhorn,rudolphschmidt}.
In terms of the familiar  Yang-Mills connection, a non-Abelian \(1\)-cohomology class is then simply a set of transition maps \(\alpha_{ij}\colon U_i\cap U_j\to\Gamma\) that compose nicely on triple overlaps (i.e.\ the triple overlap \(\mathrm d\alpha\colon U_i\cap U_j\cap U_k\to\Gamma\) is trivial) modulo gauge transformations.

\subsection{Magnetic centre symmetry of centreless Yang--Mills theory}
Given spacetime \(M\) equipped with a sufficiently fine open cover \(M=\bigcup_iU_i\),
the data of the Yang--Mills gauge field valued in the centreless gauge group \(\tilde G\) can be given in terms of \(G\)-valued (rather than \(\tilde G\)-valued) Čech cochains by
\begin{align}
    A&\in\vOmega{}^0(U;\mathrm T^*U\otimes\mathfrak g),&
    g&\in\vOmega{}^1(U;G),
\end{align}
satisfying
\begin{align}
    A_i|_{U_i\cap U_j}&=g_{ij}A_jg_{ij}^{-1}+g_{ij}\mathrm dg_{ij}^{-1}|_{U_i\cap U_j},&
    \mathrm dg&\in\vOmega{}^2(U;\operatorname Z(G)),
\end{align}
where crucially $\mathrm dg$ is valued in $Z(G)$. That is, in a \(\tilde G\)-valued Yang--Mills theory, the cocycle condition \((\mathrm d g)_{ijk}=g_{ij}g_{jk}g_{ik}^{-1}=1\) (which would describe a \(G\)-valued Yang--Mills theory) is weakened to \(g_{ij}g_{jk}g_{ik}^{-1}\in\operatorname Z(G)\).
This can be thought of as resolving \(\tilde G\) into the two-group given by the crossed module
\begin{equation}
    \left(\operatorname Z(G)\to G\right).
\end{equation}
This is a continuum non-Abelian analogue of the Villain formulation \cite{Villain:1974ir} of lattice \(\operatorname U(1)\) gauge theory (where one resolves the gauge group \(\operatorname U(1)\) into the two-group \((\mathbb Z\to\mathbb R)\)).

This Čech formulation now manifests the \(\operatorname Z(G)\)-valued  magnetic \((d-3)\)-form symmetry of \(\tilde G\)-valued Yang--Mills theory through its discrete Noether current \(\mathrm dg\), which is a \(\operatorname Z(G)\)-valued two-form\footnote{This represents a physical symmetry even though it appears to be \(\mathrm d\)-exact since there is no \(\operatorname Z(G)\)-valued one-form whose differential is \(\mathrm dg\).}. Of course, this is entirely equivalent to the usual statement that the magnetic symmetry is $\pi_1(\tilde G)$-valued.

The constraint that \(\mathrm dg\) lies in \(\operatorname Z(G)\) may be implemented using a Lagrange multiplier \(C\in\vOmega{}^{d-2}(U;\operatorname Z(G))\) as
\begin{equation}\label{eq:villain_centreless_ym}
    S = \int_M\frac12\operatorname{tr}(F\wedge\star F)+(\mathrm dg+B)\wedge C,
\end{equation}
where \(B\in\vOmega{}{}^2(U;\operatorname Z(G))\) is an auxiliary field, and \(F=\mathrm dA+\frac12[A,A]\in\Omega^2(M;\mathfrak g)\) is the usual Yang--Mills field strength (which is a true differential form unlike \(A\)), and
where in the second term \((\mathrm dg+B)\wedge C\), the \(\wedge\) actually denotes the cup product of \(\operatorname Z(G)\)-valued cocycles using the ring structure\footnote{
    For a simply connected compact simple Lie group \(G\), the centre \(\operatorname Z(G)\) is always a cyclic group or a product of two cyclic groups and hence admits a ring structure \(\mathbb Z_q\) or \(\mathbb Z_q\times\mathbb Z_{q'}\) for some \(q,q'\in\mathbb Z^+\). Here, \(q=1\) for \(\operatorname F_4\), \(\operatorname G_2\), \(\operatorname E_8\); and \(q=2\) for \(\operatorname{Spin}(2n+1)\), \(\operatorname{Sp}(n)\), \(\operatorname E_7\); and \(q=3\) for \(\operatorname E_6\); and \(q=4\) for \(\operatorname{Spin}(4n+2)\); and \((q,q')=(2,2)\) for \(\operatorname{Spin}(4n)\); and \(q=n\) for \(\operatorname{SU}(n)\).
} on \(\operatorname Z(G)\).

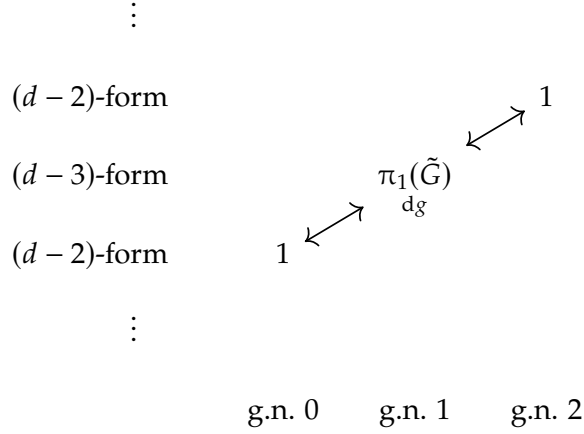
\begin{figure}
\begin{equation*}
    \begin{tikzcd}[row sep={\pgfkeysvalueof{/tikz/commutative diagrams/row sep/large},between origins}, column sep={4.5em,between origins}]
    \qquad\quad\vdots\\
    \text{\((d-2)\)-form}&&&1\dlar[leftrightarrow]\\
    \text{\((d-3)\)-form}&&\underset{\mathrm dg}{\pi_1(\tilde G)}\dlar[leftrightarrow]\\
    \text{\((d-2)\)-form}&1\\
    \qquad\quad\vdots\\
    &\text{g.n.\ \(0\)}&\text{g.n.\ \(1\)}&\text{g.n.\ \(2\)}
    \end{tikzcd}
\end{equation*}
\caption{Ghostly higher-form symmetries of \(d\)-dimensional \(\tilde G\)-valued Yang--Mills theory, where \(\tilde G\) is a centreless compact Lie group}\label{fig:magnetic_centre}
\end{figure}

The magnetic \((d-3)\)-form symmetry does not beget any descendants since there is no field whose  Batalin--Vilkovisky differential equals \(\mathrm dg\).  Hence the higher-form symmetries of \(\tilde G\)-valued Yang--Mills theory are as give in \cref{fig:magnetic_centre}.

\subsection{Electric centre symmetry of simply connected Yang--Mills theory and its descendants}

Let us now examine the electric symmetry of the simply connected \(G\)-valued Yang--Mills theory.
In this case we consider the simply connected \(G\)-valued Yang--Mills theory itself with , but where the closure of Čech cochain $g\in\vOmega{}^1(U;G)$ is now  relaxed  in the sense that   \(\mathrm dg =\mathrm d b \), where \(b\in\vOmega{}^1(U;\operatorname Z(G))\). It is exact in the sense that it is the differential of a $Z(G)$-valued 1-cocycle. 
This can be accomplished by slightly modifying \eqref{eq:villain_centreless_ym} so that, instead of \(B\in\vOmega{}^2(U;\operatorname Z(G))\), one has the differential of an auxiliary field \(b\in\vOmega{}^1(U;\operatorname Z(G))\) as in
\begin{equation}\label{eq:villain_ym}
    S = \int_M\frac14\operatorname{tr}(F\wedge\star F)+(\mathrm dg-\mathrm db)\wedge C.
\end{equation}
Now, \(\mathrm dg\) no longer represents a physical symmetry since it is exact while \(C\) is now closed on shell and hence represents a physical \(\operatorname Z(G)\)-valued one-form symmetry, namely the electric centre symmetry of simply connected Yang--Mills theory.

\begin{figure}
\begin{equation*}
    \begin{tikzcd}[row sep={\pgfkeysvalueof{/tikz/commutative diagrams/row sep/large},between origins}, column sep={4.5em,between origins}]
    \text{\(\hphantom{(\;-0)}d\)-form}&&&&&&1\dlar[leftrightarrow]\\
    \text{\((d-1)\)-form}&&&&&\underset\beta{\operatorname Z(G)}\dlar[leftrightarrow]\\
    \text{\((d-2)\)-form}&&&&\xcancel b\\
    \qquad\quad\vdots\\
    \text{\(\hphantom{()-d}2\)-form}&&&&&1\dlar[leftrightarrow]\\
    \text{\(\hphantom{()-d}1\)-form}&&&&\underset C{\operatorname Z(G)}\dlar[leftrightarrow]\\
    \text{\(\hphantom{()-d}0\)-form}&&&\underset{b^+}{\operatorname Z(G)}\dlar[leftrightarrow]\\
    \text{\(\hphantom{\;d\;}(-1)\)-form}&&\underset{\beta^+}{\operatorname Z(G)}\dlar[leftrightarrow]\\
    \text{\(\hphantom{\;d\;}(-2)\)-form}&1\\
    &\text{g.n.\ \(-3\)}&\text{g.n.\ \(-2\)}&\text{g.n.\ \(-1\)}&\text{g.n.\ \(0\)}&\text{g.n.\ \(1\)}&\text{g.n.\ \(2\)}
    \end{tikzcd}
\end{equation*}
\caption{Ghostly higher-form symmetries of \(d\)-dimensional \(G\)-valued Yang--Mills theory, where \(G\) is a simply connected compact Lie group. Note that this figure and \cref{fig:magnetic_centre} combined mimic the symmetries of Maxwell theory given in \cref{fig:symmetries_maxwell}.}
\label{fig:electric_centre}
\end{figure}
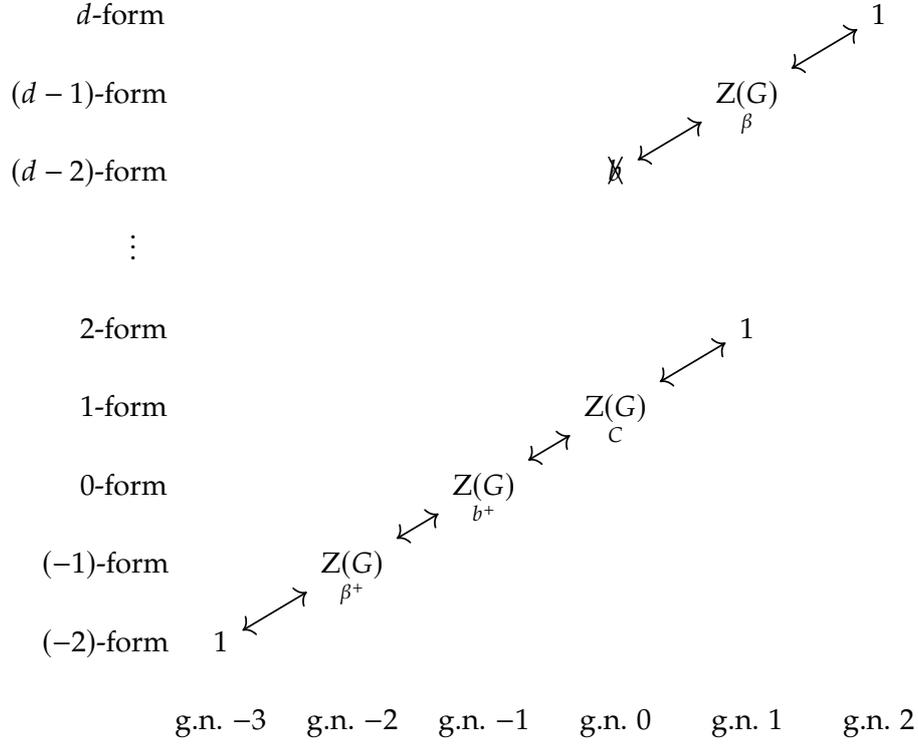

Furthermore, the theory now gains more gauge symmetries since \eqref{eq:villain_ym} is invariant up to a total derivative term when \(b\) is shifted by exact \(\operatorname Z(G)\)-valued differential forms.\footnote{
    There is no corresponding gauge symmetry shifting \(C\) by an exact \(\operatorname Z(G)\)-valued differential form
    since \(\mathrm dg\) (despite the notation) is not actually a closed \(\operatorname Z(G)\)-valued differential form off shell; \(\mathrm dg\) is only a  \(\operatorname Z(G)\)-valued differential form on shell.
}
Thus, similar to Abelian higher gauge theory, we introduce the following ghost fields and their antifields:
\begin{equation}
\begin{aligned}
    C&\in\vOmega{}^{d-2}(U;\operatorname Z(G)),&
    C^+&\in\vOmega{}^2(U;\operatorname Z(G)),\\
    \beta&\in\vOmega{}^0(U;\operatorname Z(G)),&
    \beta^+&\in\vOmega{}^d(U;\operatorname Z(G)),
\end{aligned}
\end{equation}
where \(\beta\) carries ghost number \(1\) and \(\beta^+\) carries ghost number \(-2\).
Effectively, \(b\) now serves as the `centre part' of the \(G\)-valued \(g\).

Then the corresponding Batalin--Vilkovisky action is 
\begin{equation}\label{eq:villain_ym_bv}
    S = \int_M\frac14\operatorname{tr}(F\wedge\star F)+(\mathrm dg-\mathrm db)\wedge C
    +b^+\wedge\mathrm d\beta
    +\mathcal O( c),
\end{equation}
where we have omitted terms corresponding to the gauge transformation of \(A\) and \(g\) etc.~with respect to the usual Yang--Mills ghost \(c\), which are not important here.
From \eqref{eq:villain_ym_bv}, we may read off the relevant set of Batalin--Vilkovisky differentials as
\begin{equation}
\begin{aligned}
    Qb&=\mathrm d\beta,&Q\beta&=0,&QC&=0,\\
    Qb^+&=\mathrm dC,&Q\beta^+&=\mathrm db^+,&QC^+&=\mathrm dg-\mathrm db.
\end{aligned}
\end{equation}
Now, descending from the conserved current \(J_\mathrm e\coloneqq C\), we see that
\begin{align}
    0&=QJ_\mathrm e,&
    \mathrm dJ_\mathrm e&=Qb^+,&
    \mathrm db^+&=Q\beta^+,&
    \mathrm d\beta^+&=0.
\end{align}
In addition \(\beta\) is the Noether current for a ghostly \((d-1)\)-form symmetry, but the descent chain stops after one rung:
\begin{align}
    0&=Q\beta,&
    \mathrm d\beta&=Qb,&
    Qb&\not\in\operatorname{im}_{\mathcal A}(\mathrm d).
\end{align}
Thus, the ghostly higher symmetries of \(G\)-valued Yang--Mills theory are as shown in \cref{fig:electric_centre}.

\paragraph{Wilson operators.}
As is well known, the electric and magnetic centre symmetries act on Wilson loops and 't~Hooft loops, respectively. From the action \eqref{eq:villain_ym_bv}, we see that the Wilson hypersurface \(\int_{\Sigma_{d-1}}b^+\) transforms under the ghostly \((d-1)\)-form symmetry with current \(\beta\), while the Wilson point \(\int_{\Sigma_0}\beta\) transforms under the ghostly zero-form symmetry with current \(b^+\).
The zero-form ghost symmetry with current \(b^+\) counts (modulo the periodicity of \(\operatorname Z(G)\)) the number of excitations of \(b^+\) in the Batalin--Vilkovisky-extended Hilbert space.

\subsubsection{Descendants of the centre symmetry under Higgsing}\label{ssec:higgs}
We now add to the pure \(G\)-valued Yang--Mills theory a minimally coupled scalar matter field valued in the adjoint representation \(\mathfrak g\) that acquires a vacuum expectation value. Then \(G\)  generically breaks down to a Cartan subgroup.
We will see that the infrared theory of Cartan-subgroup-valued Abelian gauge fields with massive W-bosons and 't~Hooft--Polyakov monopoles has the same higher-form symmetries as the ultraviolet theory, including the ghostly symmetries. (The case of non-ghostly higher-form symmetries is, of course, well known \cite{Bhardwaj:2023kri}).

For definiteness, let us take \(G=\operatorname{SU}(2)\),
so that the Cartan subgroup is \(\operatorname U(1)\subset\operatorname{SU}(2)\).
We must determine the charges of the particles in the infrared theory that breaks the \(\operatorname U(1)^{[1,0]},\operatorname U(1)^{[0,-1]},\operatorname U(1)^{[-1,-2]}\) electric, \(\operatorname U(1)^{[d-3,0]}\) magnetic and \(\operatorname U(1)^{[d-1,1]}\) ghostly symmetries shown in \cref{fig:symmetries_maxwell}.

In that case, the infrared theory contains massive W-bosons of electric charge \(\pm2\), so that
the electric one-form symmetry \(\operatorname U(1)^{[1,0]}\) and its descendant \(\operatorname U(1)^{[0,-1]}\)
break down to \(\mathbb Z_2^{[1,0]},\mathbb Z_2^{[0,-1]},\mathbb Z_2^{[-1,-2]}\) (according to \cref{fig:symmetries_maxwell_matter}).
Furthermore, since
\begin{equation}
    \pi_2\mleft(\operatorname{SU}(2)/\operatorname U(1)\mright)
    =\pi_2(\mathbb S^2)
    =\mathbb Z,
\end{equation}
the theory contains a 't~Hooft--Polyakov monopole \((d-4)\)-branes of magnetic charge \(\pm1\), which totally breaks the \(\operatorname U(1)^{[d-3,0]}\) magnetic symmetry.

Finally, in the infrared, the broken (\((\operatorname{SU}(2)/\operatorname U(1)\)) components of \(c\) transform nontrivially under the \(\operatorname U(1)\) Cartan part of \(c\), so that the would-be \(\operatorname U(1)^{[d-1,1]}\) and \(\operatorname U(1)^{[-1,-2]}\) (generated by \(c\) and \(c^+\) respectively) found in \cref{fig:symmetries_maxwell_matter} break down into the centre \(\mathbb Z_2^{[d-1,1]}\) and \(\mathbb Z_2^{[-1,-2]}\) respectively. Thus the symmetry of the infrared theory agrees with that of the ultraviolet theory given in \eqref{fig:electric_centre}.

\section{Higher ghost symmetry in  higher gauge theory}\label{sec:higher_gauge}
Higher gauge theory is a higher-form generalisation (categorification) of Yang--Mills theory, where the gauge symmetry is described by a Lie 2-group (or, more generally, a Lie \(\infty\)-group); see  \cite{Baez:2010ya, Borsten:2024gox} for reviews.
Whereas Yang--Mills theory has a non-Abelian one-form potential and two-form field strength, higher gauge theory in general has potentials and field strengths of higher form degree.
There exists a higher-form centre symmetry for higher gauge theory \cite{Borsten:2025diy} generalising that of Yang--Mills theory. However, in the higher-gauge-theory case we expect a richer pattern of descendants due to the fact that there exist second-order gauge transformations.

\subsection{Lightning review of adjusted higher gauge theory}
Whilst in the Abelian case any arbitrary Abelian gauge group can be used for higher gauge theory (as in \cref{ssec:higher_abelian_gauge_theory}), in the non-Abelian case, to ensure rich dynamics, there are restrictions on the structure of a 2-group that can be used: namely, those that admit a structure of \emph{adjustment} \cite{Sati:2008eg,Sati:2009ic,Samann:2019eei,Schmidt:2019pks,Kim:2019owc,Borsten:2021ljb,Tellez-Dominguez:2023wwr,Fischer:2024vak,Gagliardo:2025oio} (see \cite{Borsten:2024gox} for a review).
Concretely, we consider an adjusted higher gauge theory associated to an adjusted strict 2-group \(\mathcal G\). This class of higher gauge theories is treated in detail in \cite{Rist:2022hci,Borsten:2024gox} with explicit examples and applications.

\paragraph{Gauge 2-group.}
An adjusted strict Lie 2-group \(\mathcal G=(G,H,s,\rho,\kappa)\) is given by the data of two Lie groups \(G\) and \(H\) (with associated Lie algebras \(\mathfrak g\) and \(\mathfrak h\) respectively), a smooth group homomorphism \(s\colon H\to G\) (with infinitesimalisation \(s_{\mathfrak h}\colon\mathfrak h\to\mathfrak g\)), a smooth action \(\rho\colon G\to\operatorname{Aut}(H)\) (with infinitesimalisations \(\rho_{\mathfrak h}\colon G\to\operatorname{Aut}(\mathfrak h)\) and \(\rho_{\mathfrak g}\colon\mathfrak g\to\mathfrak{der}(H)\)), and a map \(\kappa\colon G\times \mathfrak{g} \to\mathfrak h\) (the adjustment, with infinitesimalisation \(\kappa_{\mathfrak g}\colon\mathfrak g\times\mathfrak g\to\mathfrak h\))
satisfying the axioms
\begin{equation}
\begin{gathered}
\begin{aligned}
s(\rho(g)h)&=h s(h) h^{-1}, &  \rho (s(h)) h'&=h h' h^{-1}, &
\kappa(s(h), x) & =h(\rho(x)h^{-1}),
\end{aligned}\\
\kappa(g'g, x) =\rho_{\mathfrak h}(g') \kappa(g, x)+\kappa(g', gxg^{-1}-s_{\mathfrak h}(\kappa(g, x)))
\end{gathered}
\end{equation}
for all \(h, h'\in H\) and \(g,g'\in G\) and \(x \in \mathfrak g\). For examples of adjusted Lie 2-groups, one can take \(\mathcal G\) to be the loop model of the string Lie 2-group \cite{Saemann:2017rjm,Schmidt:2019pks,Kim:2019owc}, where both \(G\) and \(H\) are non-Abelian.

\paragraph{Principal 2-bundle.}
Just as the ordinary Yang--Mills theory with gauge group \(G\) involves a principal \(G\)-bundle,
adjusted higher gauge theory with gauge 2-group \(\mathcal G\) involves a principal \(\mathcal G\)-bundle, as defined in 
\cite{Aschieri:2003mw,wockel11,Nikolaus:2011ag,1608.00401}. The adjustment \(\kappa\) does not enter into the (topological) principal 2-bundle itself, only in defining the gauge fields (connections).

Covering spacetime \(M\) by a sufficiently fine open cover \(\{U_i\}\), the data of a principal \(\mathcal G\)-bundle may be given using Čech cochains in terms of
\begin{align}
    g&\in\vOmega{}^1(U;G),&h&\in\vOmega{}^2(U;H).
\end{align}
In addition, for the dynamical theory, there exists another field \(\Lambda\) only defined on pairwise intersection of patches \cite[(3.34a)]{Borsten:2024gox}:
\begin{equation}
\Lambda\in\vOmega{}^1(U,\mathrm T^*U\otimes\mathfrak h).
\end{equation}
These obey the cocycle conditions \cite[(3.30), (3.34b)]{Borsten:2024gox}
\begin{align}\label{eq:cocycle_cond}
    0&=(\mathrm d_hg)_{ijk}\coloneq s(h_{ijk})g_{ij}g_{jk}g_{ki},\\
    0&=(\mathrm d_gh)_{ijkl}\coloneqq h_{ijl}^{-1}h_{ikl}h_{ijk}\rho_{g_{ij}}(h_{jkl}^{-1}),\\
    0&=(\mathrm d_{g,h}\Lambda)_{ik}\coloneqq \Lambda_{jk}-\Lambda_{ik}+\rho(g^{-1}_{jk})\Lambda_{ij}-\rho(g^{-1}_{ik})(h_{ijk}(\mathrm d+A_i)h^{-1}_{ijk}).
\end{align}

\paragraph{Potentials.}
The higher gauge theory with higher gauge symmetry \(\mathcal G\) has locally defined potentials
\begin{align}
B&\in \Omega^2(M, \mathfrak h),&A&\in \Omega^1(M, \mathfrak g),
\end{align}
with corresponding (globally defined) field strengths
\begin{equation}
\begin{aligned}
    H&\coloneqq\mathrm dB+[A,B]-\kappa_{\mathfrak g}(A, F)\in\Omega^3(M,\mathfrak h),\\
    F&\coloneqq\mathrm dA+\frac12[A,A]+s_{\mathfrak h}(B)\in\Omega^2(M,\mathfrak g).
\end{aligned}
\end{equation}
The fields transform under (finite) gauge transformations given by  \cite[(3.35),  (3.36)]{Borsten:2024gox}
\begin{equation}\label{eq:adjusted_gauge_transform_finite}
\begin{aligned}
A & {\overset{(\hat g,\lambda)}\mapsto}\tilde  g^{-1} A\tilde  g+\hat g^{-1}\mathrm d\hat g-s_{\mathfrak h}(\lambda),\\
B & {\overset{(\hat g,\lambda)}\mapsto} \rho_{\mathfrak h}(\hat g^{-1})B +\mathrm{d} \lambda+\left[\hat g^{-1} A\tilde  g+\hat g^{-1}\mathrm d\hat g-s_{\mathfrak h}(\lambda),  \lambda\right]+\frac{1}{2}\left[\lambda, \lambda\right]-\kappa\mleft(\hat g^{-1}, F\mright),\\
\lambda&{\overset{(\hat g,\hat h)}\mapsto}
\lambda+\rho(\hat g^{-1})(\hat h^{-1}(\mathrm d+\rho_{\mathfrak g}(A))\hat h),
\end{aligned}
\end{equation}
where the gauge parameters are \(\hat g\colon M\to G\) and \(\lambda \in \Omega^1(M,\mathfrak h)\), together with a second-order gauge parameter \(\hat h\colon M\to H\) under which \(\lambda\) transforms.
The gauge parameters \((\hat g,\lambda,\hat h)\) correspond respectively to the kinematic data \((g,\Lambda,h)\).

Since the field strengths now transform covariantly, we may write down an action
\begin{equation}
    S = \int \frac12H\wedge\star H+\frac12F\wedge\star F
\end{equation}
using suitable invariant inner products.

\subsection{Centre symmetries and their descendants in higher gauge theory}
\begin{figure}
\begin{equation*}
    \begin{tikzcd}[row sep={\pgfkeysvalueof{/tikz/commutative diagrams/row sep/large},between origins}, column sep={4.5em,between origins}]
    \text{\((d+1)\)-form}&&&&&&1\dlar[leftrightarrow]\\
    \text{\(\hphantom{(\;-0)}d\)-form}&&&&&\underset{\beta_{h,2}}{Z_H}\dlar[leftrightarrow]\dlar[leftrightarrow]&1\dlar[leftrightarrow]\\
    \text{\((d-1)\)-form}&&&&\underset{\beta_{h,1}}{Z_H}\dlar[leftrightarrow]&\underset\beta{Z_G}\dlar[leftrightarrow]&1\dlar[leftrightarrow]\\
    \text{\((d-2)\)-form}&&&\xcancel{b_h}&\xcancel{b_g}&\underset{\beta_\Lambda}{Z_{\mathfrak h}}\dlar[leftrightarrow]\\
    \text{\((d-3)\)-form}&&&&\xcancel{b_\Lambda}\\
    \qquad\quad\vdots\\
    \text{\(\hphantom{()-d}3\)-form}&&&&&1\dlar[leftrightarrow]\\
    \text{\(\hphantom{()-d}2\)-form}&&&&\underset{C_\Lambda}{Z_{\mathfrak h}}\dlar[leftrightarrow]&1\dlar[leftrightarrow]&1\dlar[leftrightarrow]\\
    \text{\(\hphantom{()-d}1\)-form}&&&\underset{b_\Lambda^+}{Z_{\mathfrak h}}\dlar[leftrightarrow]&\underset{C_g}{Z_G}\dlar[leftrightarrow]&\underset{C_h}{Z_H}\dlar[leftrightarrow]\\
    \text{\(\hphantom{()-d}0\)-form}&&\underset{\beta_\Lambda^+}{Z_{\mathfrak h}}\dlar[leftrightarrow]&\underset{b_g^+}{Z_G}\dlar[leftrightarrow]&\underset{b^+_h}{Z_H}\dlar[leftrightarrow]\\
    \text{\(\hphantom{\;d\;}(-1)\)-form}&1&\underset{\beta_g^+}{Z_G}\dlar[leftrightarrow]&\underset{\beta_{h,1}^+}{Z_H}\dlar[leftrightarrow]\\
    \text{\(\hphantom{\;d\;}(-2)\)-form}&1&\underset{\beta_{h,2}^+}{Z_H}\dlar[leftrightarrow]\\
    \text{\(\hphantom{\;d\;}(-3)\)-form}&1\\
    &\text{g.n.\ \(-3\)}&\text{g.n.\ \(-2\)}&\text{g.n.\ \(-1\)}&\text{g.n.\ \(0\)}&\text{g.n.\ \(1\)}&\text{g.n.\ \(2\)}
    \end{tikzcd}
\end{equation*}
\caption{Ghostly higher-form symmetries of adjusted higher gauge theory}\label{fig:adjusted_symmetries}
\end{figure}
We now consider ghostly analogues of the centre symmetry of Yang--Mills theory as discussed in \cref{sec:centre}.
The non-ghostly subset of these symmetries was previously discussed in \cite{Borsten:2025diy}; here we complete this discussion by also considering ghostly symmetries as well.

Recall (cf.\ \cite{Borsten:2025diy}) that the centre symmetry of Yang--Mills theory may be detected by considering a spacetime of the form \(M=\Sigma\times\mathbb S^1\) (i.e.\ with periodic Euclidean time, or equivalent at finite temperature) and looking at non-periodic `gauge transformations'\footnote{These are, of course, not true gauge transformations since (1) they are not periodic and (2) they act nontrivially on physical observables such as the Polyakov loop.} that nevertheless preserve the action;
the same procedure applies to adjusted higher gauge theory. 
In previous work \cite{Borsten:2025diy}
we examined the gauge transformations for \(A\) and \(B\) in \eqref{eq:adjusted_gauge_transform_finite}, to obtain the one-form and two-form centre symmetries \(Z_G^{[1,0]}\) and \(Z_{\mathfrak h}^{[2,0]}\), where
\begin{align}
    Z_G&=\left\{\gamma\in\operatorname Z(G)\middle|\kappa(\gamma^{-1},\mathfrak g)=0,\;\rho(\gamma)=\operatorname{id}_{\mathfrak h}\right\},&
    Z_{\mathfrak h}&=\ker(s)\cap\operatorname Z(\mathfrak h).
\end{align}
In higher gauge theory, however, the ghost field \(\lambda\) also undergoes higher-order gauge transformations in \eqref{eq:adjusted_gauge_transform_finite}, so that by the same analysis we obtain a ghostly one-form centre symmetry \(Z_H^{[1,1]}\), where
\begin{equation}
    Z_H=\left\{
    h\in H\middle|
    h\rho_{\mathfrak g}(x)h^{-1}
    =1_H\;\forall x\in\mathfrak g
    \right\}.
\end{equation}
In order to investigate their descendants, we work in the Čech formalism of \cref{sec:centre}, so that the conditions \eqref{eq:cocycle_cond} are relaxed to
\begin{align}
    \mathrm d_hg&=\mathrm db_g,&
    \mathrm d_gh&=\mathrm db_h,&
    \mathrm d_{g,h}\Lambda&=\mathrm db_\Lambda,&    
\end{align}
where \(b_g\in\vOmega{}^2(U;Z_G)\) and \(b_h\in\vOmega{}^3(U;Z_H)\)
and \(b_\Lambda\in\vOmega{}^1(U;\mathrm T^*U\otimes Z_{\mathfrak h})\) serve to parameterise the `centre part' of \(g\in\vOmega{}^2(U;G)\) and \(h\in\vOmega{}^2(U;H)\) and \(\Lambda\in\vOmega{}^1(U;\mathrm T^*U\otimes\mathfrak h)\), so that one adds to the action Lagrange-multiplier terms
\[
    S = \int_M \dotsb + (\mathrm d_hg-\mathrm db_g)\wedge C_g
    + (\mathrm d_gh-\mathrm db_h)\wedge C_h
    + (\mathrm d_{g,h}\Lambda-\mathrm db_\Lambda)\wedge C_\Lambda
\]
with Lagrange-multiplier fields \(C_g\in\vOmega{}^{d-3}(U;Z_G)\) and \(C_h\in\vOmega{}^{d-4}(U;Z_H)\) and \(C_\Lambda\in\vOmega{}^{d-3}(U;\mathrm TU\otimes Z_H)\),
similar to the Yang--Mills case in \cref{sec:centre}, and one adds ghost fields \(\beta_g\), \(\beta_{1,h}\) (with higher-order ghost \(\beta_{2,h}\)) and \(\beta_\Lambda\) for the gauge transformations of \(b_g\), \(b_h\), and \(b_\Lambda\) respectively.

Then proceeding analogously as for the discussion of Yang--Mills theory in \eqref{sec:centre}, one obtains the ghostly higher-form symmetries of adjusted higher gauge theory shown in \cref{fig:adjusted_symmetries}.

\newpage

\bibliographystyle{JHEP}
\bibliography{biblio}

\end{document}